\documentclass[preprint,12pt]{aastex}

\usepackage{emulateapj5}

\slugcomment{Accepted to ApJ v594 September 2003}

\shorttitle{Optical Spectra of T Dwarfs}
\shortauthors{Burgasser et al.}

\begin{document}

\title{The Spectra of T Dwarfs. II. Red Optical Data}

\author{
Adam J.\ Burgasser\altaffilmark{1,2},
J.\ Davy Kirkpatrick\altaffilmark{3},
James Liebert\altaffilmark{4},
and
Adam Burrows\altaffilmark{4}
}

\altaffiltext{1}{Department of Astronomy \& Astrophysics,
University of California
at Los Angeles, Los Angeles,
CA, 90095-1562; adam@astro.ucla.edu}
\altaffiltext{2}{Hubble Fellow}
\altaffiltext{3}{Infrared Processing and Analysis Center, M/S 100-22,
California Institute of Technology, Pasadena, CA 91125; davy@ipac.caltech.edu}
\altaffiltext{4}{Steward Observatory, University of Arizona,
Tucson, AZ 85721; liebert@as.arizona.edu, burrows@as.arizona.edu}

\begin{abstract}
We present 6300--10100 {\AA} spectra for a sample of 13 T dwarfs
observed using LRIS mounted on the Keck I 10m Telescope.
A variety of features
are identified and analyzed, including pressure-broadened
K I and Na I doublets; narrow Cs I and Rb I lines; weak CaH, CrH, and FeH
bands; strong H$_2$O absorption; and a possible weak CH$_4$ band.  H$\alpha$
emission is detected in three of the T dwarfs, strong in the
previously reported active T dwarf 2MASS 1237+6526 and weak
in SDSS 1254$-$0122 and 2MASS 1047+2124.  None of the T dwarfs exhibit
Li I absorption.  Guided by the evolution of optical
spectral features with near-infrared spectral type, we derive a
parallel optical classification scheme, focusing on spectral types
T5 to T8, anchored to select spectral standards.
We find general agreement
between optical and near-infrared types for nearly all of the
T dwarfs so far observed, including two earlier-type T dwarfs, within
our classification uncertainties ($\sim$1 subtype).
These results suggest that competing
gravity and temperature effects compensate for each other
over the 0.6--2.5 $\micron$ spectral region.  We identify
one possible means of disentangling these effects
by comparing the strength of the K I red wing to the 9250 {\AA}
H$_2$O band.
One of our objects, 2MASS 0937+2931, exhibits a peculiar spectrum, with a
substantial red slope and relatively strong FeH absorption, both consequences
of a metal-deficient atmosphere.
Based on its near-infrared properties and substantial space motion,
this object may be a thick disk or halo brown dwarf.
\end{abstract}

\keywords{stars: activity ---
stars: fundamental parameters ---
stars: individual (2MASS J09373487+2931409)
stars: low mass, brown dwarfs
}

\section{Introduction}
Current classification schemes for the two latest spectral classes,
L dwarfs \citep{kir99,mrt99} and T dwarfs
\citep{me02a,geb02}, are defined in different spectral regions. L dwarf
classification is tied to the red optical
region (6300--10000 {\AA}), enabling straightforward comparison with
M dwarf classification schemes \citep{kir91}, and hence a well-defined
demarcation of the M and L spectral classes.
T dwarf classification is tied to the near-infrared region
(1--2.5 $\micron$), taking advantage of both greater
relative brightness ($\sim$ 75\% of the emergent flux)
and the presence of strong H$_2$O and defining CH$_4$
bands.
Classifications in different spectral regions cannot be expected to be
identical, however, due to the competing influences of temperature,
gravity, and composition at different wavelengths.
This is particularly the case for
late-type M, L, and T dwarfs, whose molecular-rich atmospheres
also contain condensate clouds,
predominantly affecting near-infrared spectral flux
\citep{tsu99,ack01}.
The influence of these chemical and physical variables
muddle the characterization of the L/T transition,
a spectromorphological boundary that encompasses a dramatic
shift in near-infrared colors (J--K $\approx$ 2 $\rightarrow$ 0),
condensate opacity,
and overall spectral energy distribution.
Clearly, the evolution of both near-infrared and optical spectra
must be examined to fully characterize this critical transition.

The red optical spectra of T dwarfs are of interest in their own right,
as they encompass unique features and
diagnostics of physical parameters.
Early investigation of the prototype of this class,
Gliese 229B \citep{nak95,opp95}, revealed strong H$_2$O and weak CH$_4$
molecular absorption (counterparts to the strong bands that dominate the
near-infrared spectrum), Cs I alkali lines, and
a steep red spectral slope shortward of 1
$\micron$ \citep{opp98,sch98}.  The latter feature, attributable to
the pressure-broadened wings of the 7665/7699 {\AA} K I
resonant doublet lines \citep{tsu99,lie00,bur00}, is responsible for
the extremely red optical/near-infrared colors of this object (R--J $\sim$ 9.1;
Matthews et al.\ 1996, Golimowski et al.\ 1998), a property that
has been exploited in the search for T dwarfs in wide-field
surveys such as the Two Micron All Sky Survey \citep[hereafter 2MASS]{skr97}
and the Sloan Digital Sky Survey \citep[hereafter SDSS]{yor00}.
\citet{bur02} have shown that these heavily-broadened lines and their 5890/5896
{\AA} Na I counterparts are sensitive to
both effective temperature (T$_{eff}$) and specific gravity.  The much weaker Cs I lines
are key tracers of
atmospheric rainout \citep{gri99,lod99}.
Additional molecular bands of FeH and CrH reported in the
spectra of mid-type T dwarfs \citep{me00b} are possible tracers of
condensate cloud opacity across the L/T transition \citep{me02c}.
Finally, persistent
H$\alpha$ emission detected in the T dwarf 2MASS
1237+6526 \citep{me00b,me02b}, whose atmosphere is predicted to be too
cool to support a substantial chromosphere \citep{moh02,gel02},
is not yet understood and has been the subject of
some debate \citep{hal02a,hal02b,lie03}.

In this article, we continue the spectroscopic investigation of
T dwarfs initiated in \citet[Paper I]{me02a} by examining their red
optical spectra.  This article complements that of \citet{kir03},
which discusses the optical L/T transition in detail.
In $\S$2 we describe the acquisition and
reduction of the data, obtained with the Low
Resolution Imaging Spectrograph \citep[hereafter LRIS]{oke95}
mounted on the Keck I 10m Telescope. In
$\S$3 we present the reduced spectra, analyzing
in detail the spectral features observed.
In $\S$4 we present a classification scheme for mid- and late-type
T dwarfs using red optical
spectra that is parallel to, but independent of, the near-infrared
scheme of \citet{me02a}.
In $\S$5 we discuss our results, including
the physical properties of the peculiar T dwarf 2MASS 0937+2931,
a possible means of disentangling gravity and temperature in red optical
spectra, and
a comparison between
T dwarf optical and near-infrared classifications.
We summarize our results in $\S$6.

\section{Observations}

We acquired LRIS red optical spectra for a sample of 13 T dwarfs identified
in 2MASS \citep{me99,me00a,me00c,me02a,me03a} and
SDSS \citep{leg00} on three
nights, 2000 March 5, 2001 February 20, and 2003 January 2
(UT).  An observing log is given in Table 1.
Conditions during our March 2000 and January 2003 observations
were clear with seeing $\lesssim$ 1$\arcsec$; our
February 2001 observations were made through light clouds with 1$\arcsec$ seeing.
LRIS is nominally a dual spectrograph
with a 5500 {\AA} dichroic mirror
separating incoming light into two
(red and blue) channels.  For all observations, we used only the
red channel with a mirror bypassing the dichroic.
The OG570 order-blocking filter was employed to suppress higher-order light,
and a 1$\arcsec$ (4.7 pixels) slit was used with the 400 lines mm$^{-1}$
grating blazed at 8500 {\AA}, yielding 6300--10100 {\AA} spectra with
$\sim$7 {\AA} resolution (R $\sim$ 1200).
Dispersion on the chip was 1.9 {\AA} pixel$^{-1}$.

Targets were nominally acquired using the facility guide
camera; objects that were too optically faint to be seen in the guider
image
were placed into the slit by blind offset from a nearby visible source,
using offset values
determined from 2MASS coordinates.
Individual exposures ranged from 1200 to 1800 sec, with up to four observations
obtained on a particular night.
Observations of the B1 V flux standard
Hiltner 600 \citep{ham94}
were obtained for flux calibration on all runs
using the same instrumental configuration.
To correct for telluric absorption,
we observed either DA/DC/DZ white dwarfs selected from \citet{mco99}
or late-F/early-G stars before or after the
target observations and at similar airmass.  These stars typically have few or no
features in the red optical telluric absorption bands.
Quartz lamp flat-field exposures (reflected off of the interior dome)
were observed at the start of each run to calibrate detector response,
and NeAr arc lamp exposures were obtained immediately after the target
observations for wavelength calibration.

Science data were reduced using
standard IRAF\footnote{Image Reduction and Analysis Facility (IRAF)
is distributed
by the National Optical Astronomy Observatories,
which are operated by the Association of Universities for Research
in Astronomy, Inc., under cooperative agreement with the National
Science Foundation.} routines.
First, we removed the image bias.  For the March 2000 data, this was done
by median-combining a series of
one-second bias frames,
and subtracting this image from the science and calibration data.
For the February 2001 and January 2003 data, the bias level for
each row was derived from the mean counts in the overscan region and subtracted.
Images were trimmed and divided by the
median-combined and normalized
flat-field exposures
to correct for detector response. Spectra were then extracted
using the IRAF APEXTRACT package.  Curvature of the dispersion line
was determined from either the telluric or flux
standard star exposures and used as a template for the target
extractions.
All spectral data
were optimally extracted with a cleaning algorithm to eliminate
cosmic rays and spurious pixels.
Wavelength calibration was done using the arc lamp exposures
and line identifications from the NIST atomic line
database\footnote{See \url{http://physics.nist.gov/}.}.

We then computed a telluric
correction from the telluric calibrator spectra by interpolating over
atmospheric O$_2$ ($\sim$6850--6900 {\AA} B-band, $\sim$7600--7700 {\AA} A-band) and
H$_2$O ($\sim$7150--7300, $\sim$8150--8350, and
$\sim$8950--9650 {\AA}; Stevenson 1994) bands, and then
dividing this modified spectrum by the original calibrator spectrum.
This correction spectrum was applied to both target and flux standard
data to eliminate telluric absorption from the spectrum\footnote{Telluric
corrections were applied to the flux standard for the March 2000 data only, as
the February 2001 and January 2003 data were first flux calibrated, then
corrected for telluric absorption.  Reversing the order of these corrections
has a negligible affect on the resulting science data.}.
We note that we acquired the wrong
star as the telluric calibrator
for 2MASS 0937+2931.  WD 0924+199 (a.k.a\ PG 0924+199, TON 1061) is
classified DC5 \citep{put97}, with a possible indication of H$\alpha$ absorption and an
11,000 K blackbody continuum.  Our observations show a roughly 5000 K
blackbody continuum, weak and narrow H$\alpha$ absorption,
a strong Ca II resonance triplet
(8498, 8542, \& 8662 {\AA}) and a Ba II line at 6497 {\AA}.  This object is likely a
background F/G-type dwarf, and fortuitously works fine as a telluric
calibrator\footnote{We also detected the Ca II resonance triplet
in the DZA5 WD 1225-079 (a.k.a.\ PG 1225-079, K 789-37), which is known to have strong
Ca II HK absorption at 3933 and 3968 {\AA} \citep{lie87}.
WD 0552-041 (a.k.a.\ HL 4, LHS 32), variously classified from DC9 to DZ11.5 \citep{mco99}
shows no metal lines in its red optical spectrum, consistent with a pure He atmosphere.
See \citet{har03} for additional discussion of the red optical spectra of white dwarfs.}.

Finally, flux calibration was done by correcting our
Hiltner 600 observations to the spectrophotometric data given by
\citet{ham94}, being careful to interpolate over
telluric absorption regions and the Balmer H$\alpha$ line.
This correction was applied to the T dwarf data to produce
the final calibrated spectra.  If multiple spectra were obtained on
a particular night, these were coadded with a sigma-clipping algorithm
to improve signal-to-noise.

\section{The Spectra}

Reduced spectra for the T dwarfs are shown
in Figure 1 in order of their near-infrared spectral
classifications.
Features present in the spectra are noted in Figure 1 and listed in Table 2; we
discuss these features in detail
below.

\subsection{Atomic Line Features}

The strongly pressure-broadened
Na I and K I resonance doublets
suppress most of the optical flux of T dwarfs \citep{bur02}.
The broad wings of both alkali doublets can be seen clearly in the
highest signal-to-noise spectrum of our sample, that
of 2MASS 0559$-$1404, shown in Figure 2 with
absolute flux density (calibrated using I$_c$ photometry and
parallax measurements from Dahn et al.\ 2002) plotted on a logarithmic
scale.  The peak-up in flux between these features is clearly
evident, and their shapes are consistent with heavily
pressure-broadened atomic lines \citep{bur03}.

Other alkali line features present in these spectra
include the 8521/8943 {\AA} Cs I and
7800/7948 {\AA} Rb I resonant doublets.
Examination of Figure 1 shows that the Cs I lines are clearly
present in all of the T dwarfs observed, although they
become rather weak in the latest-type objects.  Rb I lines lie very close to the core
of the strong K I doublet, and are generally only seen in our
higher signal-to-noise spectra (e.g., Figure 2).

We have measured
pseudo-equivalent widths\footnote{The presence of overlying opacity from
K I, Na I, and other species prevents the
measurement of ``true continuum'' for computing equivalent
widths; hence, the reported measurements are relative to the
local ``pseudo-continuum''.}
(pEWs) for each of the Cs I and Rb I
lines as follows: First, we subtracted
a linear fit to the local pseudo-continuum, $C(\lambda)$,
to yield the line
profile, $L(\lambda)$.  Next, we fit each line profile to a Lorentzian function
of the form
\begin{equation}
%L(\lambda) = Ae^{-\frac{1}{2}(\frac{{\lambda}-{\lambda}_c}{{\Delta}{\lambda}})^2},
L(\lambda) = A\frac{{\lambda}_c/2}{({\lambda}-{\lambda}_o)^2 + ({\lambda}_c/2)^2},
\end{equation}
where $A$, ${\lambda}_o$ (the transition
wavelength in {\AA}), and ${\lambda}_c$ (the effective line width in {\AA})
were left as free parameters.  Fits were made iteratively and confirmed by eye to eliminate
spurious matches to noise spikes.
Finally, pseudo-equivalent widths were derived from
\begin{equation}
pEW = -\frac{{\pi}A}{C({\lambda}_o)}.
\end{equation}
Uncertainties, or upper limits in the cases where no line could be visually
identified, were derived from the uncertainty in the continuum fit
and residuals in a 40{\AA} region of the spectrum around each alkali
line after the line profile fit.
In cases where the continuum was essentially undetected, no pEW upper limits
could be derived.
Values are listed in Table 3.

Early- and mid-type T dwarfs have Cs I pEWs $\approx$ 7--9 {\AA},
with the exception of 2MASS 0755+2212, whose short wavelength spectrum
is partly filled in by flux from a nearly-aligned background galaxy \citep{me02a}.
Both Cs I lines weaken in the later-type
objects, particularly the 8521 {\AA} line, which is weakest and only marginally detected
in the T8 2MASS 0415$-$0935 (1.8$\pm$1.4 {\AA}).  The residual strength of the
8943 {\AA} line may be the result of
contamination by CH$_4$ absorption ($\S$3.2; Figure 4).
These trends are consistent with chemical equilibrium calculations by \citet{lod99},
which predict a gradual chemical depletion of Cs to CsCl at T $\lesssim$ 1300 K,
accelerated at T $\lesssim$ 1000 K as NaCl (and atomic Na)
converts to Na$_2$S and liberates
elemental Cl.  Note that K is depleted to KCl around the same temperature
\citep{lod99}, weakening the K I wing and reducing the
suppression of the local pseudo-continuum around the Cs I
lines.  That the Cs I lines nevertheless wane in the later T dwarfs
provides supporting evidence that elemental Cs is in fact depleted in their photospheres.

The 7948 {\AA} Rb I line could be measured in about two-thirds of the objects
in our sample.  Line strengths for the early- and mid-type T dwarfs vary
substantially, likely due to the decreased signal-to-noise in this spectral
region.
Surprisingly, many of the latest-type T dwarfs shown detectable, albeit marginal,
7948 {\AA} Rb I absorption.
Figure 3 shows a close-up
around this line for five T dwarfs with
positive detections;
for 2MASS 0727+1710, Gliese 570D, and 2MASS 0415$-$0935, these
detections are clearly marginal.  Nonetheless, the resurgence of the Rb I
line could be a real effect.
As with K, atomic Rb is depleted to RbCl for T $\lesssim$ 1000 K
\citep{lod99}.  The lower depletion temperature for Rb implies a much higher
relative abundance than Cs in the later T dwarfs, particularly as Rb is
roughly 20 times more abundant in a Solar metallicity mixture \citep{and89}.
Furthermore, the weakening of the K I red wing is likely to enhance
Rb I pEWs.
The 7800 {\AA} Rb I line is only detected in our
two brightest T dwarfs, 2MASS 0559$-$1404 and 2MASS 1503+2525, with
pEWs of 6$\pm$5 and 12$\pm$5 {\AA}, respectively.

One glaring absence among the alkali features is the 6708 {\AA}
Li I resonance doublet.  2MASS 0559$-$1404 shows
a possible feature close to this line (Figure 2), but it is offset
at 6690 {\AA} and is likely a residual noise spike.
All of the objects in our sample
are presumably substellar based on their low T$_{eff}$s \citep{bur97,me02a},
and a good fraction likely have masses below the 0.06 M$_{\sun}$
limit for Li core fusion depletion \citep{cha97}.  However, none exhibit
Li I absorption.  Atomic Li
is chemically depleted to LiCl, LiH, and LiOH for T $\lesssim$ 1500 \citep{lod99},
at hotter temperatures than Na, K, Rb, and Cs.
A general weakening of the Li I line is observed
in substellar late-type L dwarfs \citep{kir00}, and the absence of this line
in the T dwarfs is consistent with a temperature effect.
However, we note that these observations are not consistent with
current spectral models that use
the \citet{lod99} chemical abundances, which
continue to exhibit detectable Li I absorption down to T$_{eff}$ $\approx$ 600--800 K
\citep{all01,bur02}.
Confirmation of chemical depletion for all of the alkali species
will require examination of their respective
chloride bands at far-infrared wavelengths.

\subsection{Molecular Features}

The most prominent molecular features in the red optical spectra of T dwarfs
are the H$_2$O bands at
9250 and 9450 {\AA}.  These bands
strengthen considerably toward the latest-type objects in our
sample, but are generally weaker than
their near-infrared counterparts.
As is the case in the near-infrared, the steam bands overlap
telluric H$_2$O absorption at 8950--9700 {\AA} (strongest from 9300--9650 {\AA};
Stevenson 1994); however, the hotter
brown dwarf bands clearly extend blueward to 9250 {\AA}.

The 9896 {\AA} Wing-Ford FeH band is
seen quite clearly in many of
the spectra of Figure 1; Figure 2 of \citet{me02c} shows a close-up
of this feature and the 9969
{\AA} CrH band. Both bands are
strongest in mid-type T dwarfs like
2MASS 1534$-$2952AB, but are weak or absent in the latest-type T dwarfs.
The identification of the CrH band may be in error,
as \citet{cus03} find that absorption in this
region for M and L dwarfs is entirely attributable to FeH absorption.  While opacity data from
\citet{bur02b} clearly show a CrH bandhead at this wavelength, its visibility depends
on the relative abundance of CrH to FeH, as their opacities
are roughly equal in strength at these wavelengths \citep{dul03}.
\citet{lod99} predicts an equilibrium abundance CrH/FeH $\approx$ 10$^{-3}$
for 1800 $\lesssim$ T $\lesssim$ 2500 K, but a higher relative abundance of
$\sim$0.3 for T $\lesssim$ 1500 K.  Hence, it is entirely possible
that the 9969 {\AA} CrH band is
by FeH absorption in M and L dwarf spectra, but
detectable in the cooler T dwarf spectra.  Higher-resolution observations
are required to test this possibility.
The shorter-wavelength counterparts of the FeH and CrH bands at 8692 and 8611
{\AA}, respectively, are present in the spectrum of SDSS 1254$-$0122,
but absent in the remaining spectra.

A weak signature of CaH at
6750--7150 {\AA}, seen in M and L
dwarf spectra, also appears to be present but weak in
the spectra of 2MASS 0559$-$1404 (Figure 2), SDSS 1254$-$0122, and 2MASS 1503+2525,
and is not seen in any of the other spectra.  In chemical equilibrium,
CaH and elemental Ca are expected to
be depleted to the solid minerals perovskite (CaTiO$_3$),
grossite (CaAl$_2$O$_4$), hibonite (CaAl$_{12}$O$_{19}$), and gehlenite
(Ca$_2$Al$_2$SiO$_7$) at T $\lesssim$ 1900 K \citep{lod99}.  The detection of this weak
band in mid-type T dwarfs is therefore unexpected and may be indicative of non-equilibrium
processes.  However, improved opacities for CaH are needed before this scenario
can be properly explored.

While CH$_4$ is ubiquitous in the near-infrared spectra of T dwarfs,
only the relatively weak 8950 {\AA} CH$_4$ band,
noted by \citet{opp98} in Gliese 229B, is likely to be present in the red optical
spectra of the latest-type T dwarfs.  Figure 4 shows a close-up of this
spectral region for six T dwarfs including Gliese 229B.
Absorption at the base of the 8943 {\AA}
Cs I line clearly becomes very broad in the later-type
objects, and is strongest
in 2MASS 0415$-$0935.  As shown in this figure, the breadth of this absorption is much greater
than that expected for a pressure-broadened line, which exhibits a Lorentzian line profile
at detunings of a few 10s of Angstroms \citep{bur03}.
Furthermore,
Cs I pEWs are weakening in the latest-type T dwarfs, while the broad
absorption in this region is clearly strengthening.
Below the spectra we plot a CH$_4$ opacity spectrum generated from the HITRAN database
\citep[see references in Burrows et al.\ 1997]{rot98}.
While the broad trough around the Cs I line, and a weak feature
at 8835 {\AA} in the spectrum of 2MASS 0415$-$0935, are
generally consistent with the CH$_4$ opacity data, none of the T dwarf
spectra, including that of Gliese 229B, exhibit the strongest feature in the
opacity data
at 8875 {\AA}.  However, it is important to consider that
the CH$_4$ opacity data is based on laboratory and planetary
measurements made at temperatures significantly below those of a typical T dwarf photosphere; hence, the
overall band shape may be somewhat different in the latter environment.
Pending improved opacity data, we consider
the detection of CH$_4$ at 8950 {\AA} to be tentative.
Three other CH$_4$ features at 7300, 8000, and 10000 {\AA} seen in planetary and laboratory spectra
\citep{dic77} are intrinsically weaker than the 8950 {\AA} band and
are not seen in the T dwarf data.

\subsection{H$\alpha$ Emission}

H$\alpha$ emission at 6563 {\AA}
is detected in three objects in our sample: 2MASS 1047+2124,
2MASS 1237+6526 (previously reported in Burgasser et al.\ 2000a), and
SDSS 1254$-$0122.  Figure 5 shows a close-up
of the H$\alpha$ spectral region for these three objects.  Note that the emission
lines in 2MASS 1047+2124
and SDSS 1254$-$0122 are very weak, detected at the 2.2 and 3$\sigma$ levels, respectively,
while the emission in 2MASS 1237+6526 is substantial.  None of the other T dwarfs
exhibit discernible emission at 6563 {\AA}.

Emission fluxes and 3$\sigma$ upper limits
are listed in Table 4, along with $\log{L_{H{\alpha}}/L_{bol}}$
estimates, calculated from 2MASS J-band magnitudes and using a linear
interpolation of the J-band bolometric correction as a function of
near-infrared spectral type,
\begin{equation}
BC_J = M_{bol} - M_J = 2.97 - 0.11{\times}SpT,
\end{equation}
(SpT(T5) = 5, etc.).  This relation is based
on luminosity estimates for
2MASS 0559$-$1404 \citep[$BC_J = 2.43{\pm}0.07$]{me01},
Gliese 229B \citep[$BC_J = 2.19{\pm}0.10$]{leg99},
and Gliese 570D \citep[$BC_J = 2.09{\pm}0.10$]{geb01}.  For SDSS 1254$-$0122,
we used an average $BC_J$ between 2MASS 0559$-$1404 and
the latest-type L dwarfs \citep[$BC_J \approx 1.6$]{rei01}.
Note that pEW measurements of the H$\alpha$ line
is prohibited in most of these objects by the absence of a detected continuum.

The detection of H$\alpha$ emission in three T dwarfs appears
to stand contrary
to the reduced ionization fractions predicted in their very cool atmospheres
\citep{moh02,gel02}.  The lack of ionized material prevents the
coupling of magnetic field lines to the upper atmosphere, thereby
discouraging the formation of a substantial chromosphere via collisional
heating \citep{moh02}.  However, the relative emission luminosity of 2MASS 1047+2124 and
SDSS 1254$-$0122 is roughly 2 orders of magnitude lower than that of
active mid-type M dwarfs \citep[${\langle}\log{L_{H{\alpha}}/L_{bol}}{\rangle} \approx -3.5$]{haw96},
and is substantially lower than that of active L dwarfs \citep{giz00}, as shown in Figure 3
of \citet{me02b}.  Hence, the emission in these objects is
consistent with reduced, but existent, chromospheric activity.
The strong, persistent emission in 2MASS 1237+6526 is discussed in further detail in
\citet{me02b}.

One other emission feature at 7740 {\AA} is detected in the spectrum
of 2MASS 0755+2212, noted in Figure 1.  As described in \citet{me02a}, this object is aligned
with a background galaxy, and the observed emission line is consistent with H$\alpha$
redshifted to $z = 0.18$. The background galaxy also
contributes continuum flux in the
blue portion of the spectrum of 2MASS 0755+2212, filling in some of the alkali absorption features.

\section{Classification of T dwarfs in the Red Optical}

The spectral features described above show clear trends with
near-infrared spectral type: an increasing spectral slope
and strengthening H$_2$O absorption;
a peak in the Cs I line strengths in the early-/mid-type T dwarfs;
loss of FeH and CrH absorptions at 8692 and 8611 {\AA} in the early-type
T dwarfs, and at 9896 and 9969 {\AA} in the late-type T dwarfs; and the
possible emergence of CH$_4$ at 8950 {\AA} in the latest-type T dwarfs.  Hence, the makings of
a spectral sequence in the red optical that parallels the near-infrared
sequence are apparent.

The low signal-to-noise in most of these spectra implies that a detailed
scheme is generally not possible.  We have
therefore aimed at deriving a rough classification that parallels
the near-infrared scheme, enabling a general comparison between near-infrared and optical
spectral morphologies.
We stress that our optical scheme is
based entirely on red optical features, and that we have no {\em a priori}
expectation that near-infrared and optical spectral morphologies must necessarily
coincide.
We have augmented our observed spectral data with T dwarf LRIS data from
\citet{me00b} and \citet{kir03}.
All of the spectra in our full
sample have therefore been acquired and reduced in an identical
manner (with the exception of telluric absorption correction) and can therefore
be reliably compared.

\subsection{Spectral Standards}

Our procedures for classification followed the basic tenets of the MK
method \citep{mor73}.  We chose four T dwarfs --
SDSS 1254$-$0122 (T2), 2MASS 0559$-$1404 (T5),
SDSS 1624+0029 (T6), and
2MASS 0415$-$0935 (T8) -- and the L8
2MASS 1632+1904 \citep{kir99} as our spectral standards.
Three of the T dwarfs are near-infrared standards in the
\citet{me02a} classification scheme, while SDSS 1624+0029 replaces the
T6 near-infrared standard 2MASS 1225$-$2739AB, which has been
identified as an unequal-magnitude binary \citep{me03b}.
Note that the spectra of 2MASS 1632+1904 and SDSS 1624+0029 have not been
corrected for telluric absorption.
To improve the signal-to-noise of the spectra of SDSS 1254$-$0122 and 2MASS 0415$-$0935,
we combined our data with those of \citet{kir03}.
We did not use the T1 and T3 near-infrared
standards SDSS 0837$-$0000 and SDSS 1021$-$0304 because of their low signal-to-noise data.
This choice leaves SDSS 1254$-$0122 as the only early-type T dwarf standard, so classifications
between L8 and T5 are not intended to be robust;
see \citet{kir03} for a more detailed discussion on the optical spectra
of early-type T dwarfs.  We have nevertheless left placeholders for the
unassigned intermediate subtypes (T0, T1, T3, T4, and T7)
to allow a consistent comparison with near-infrared classifications.
To avoid confusion, we hereafter designate near-infrared
and optical classifications
as T$_n$ and T$_o$, respectively.

All of the spectral standards are plotted together in Figure 6 in both linear
and logarithmic scales.  The trends discussed above are readily
apparent over this short sequence.  More importantly, the various standards are
clearly distinguishable, and hence represent distinct morphological
classes.  In other words, the selected standards would have
been chosen to represent separate optical classes regardless of their near-infrared
types.

\subsection{Classification by Visual Comparison}

Classification of the remaining spectra in our full sample was first done
by visual comparison to the spectral standards.  We have found that
the optimal means of doing this is by comparison on a
logarithmic scale (e.g., right panel in Figure 6).
This method
permits simultaneous examination of spectral slopes ranging over two
orders of magnitude in flux density with weaker molecular and atomic features.
We have made visual classifications for all objects with LRIS spectra previously
typed T$_n$0 or later; values are listed in Table 7.

In general, visual types for the T dwarfs correspond well
with the near-infrared classifications, agreeing to within $\pm$1 subtype
for nearly every source.
One exception is SDSS 0423$-$0414, which has an earlier-type
spectrum than our L8 standard; \citet{kir03},
who classify this object as L7.5 in the optical, discusses this mismatch
and its possible physical basis in detail.  2MASS 0937+2931 is assigned a peculiar visual
classification because of its extreme red spectral slope, far in excess of any of the standards.
We discuss this object
in further detail in $\S$5.2.

\subsection{Classification by Spectral Indices}

A more quantitative approach to classification is through the use of spectral indices,
ratios of flux
or flux density that measure the strengths of particular absorption or
pseudo-continuum features.
We examined the behavior of 20 indices sampling the Cs I, FeH, CrH, H$_2$O, and spectral
slope features on our T dwarf spectra.
Some of these were taken from the literature and possibly modified \citep{kir99,mrt99,geb02,bur02},
and some are of our own construction.
The most useful indices are those that show the clearest trends and greatest contrast
with spectral type in the standards.  All of the indices were compared, and
those that best satisfied these criteria were chosen.

Eight promising indices were identified, defined in Table 5.
The Cs I indices are essentially identical to those defined by \citet{kir99}, measuring the strengths
of the 8521 and 8943 {\AA} Cs I lines, respectively.  These indices appear to peak around T$_o$2,
then weaken toward the later subclasses, consistent with the behavior of the corresponding pEWs.
The H$_2$O index samples the 9250 {\AA} band and may be weakly affected
by telluric absorption, although it has been defined to avoid the strongest absorption
longward of 9300 {\AA} \citep{ste94}.  This index increases dramatically beyond roughly T$_o$2.
The FeH(A) and CrH(A) indices measure the
shorter wavelength bands of these molecules,
which disappear around type T$_o$2, after which the
indices essentially measure a limited range of spectral slope.  The FeH(B) and CrH(B) indices measure the longer
wavelength bands, and show rather complex behavior, weakening in the late L and early
T dwarfs, strengthening again in the mid-type T dwarfs, and finally weakening in the
late T dwarfs. \citet{me02c} have proposed that this trend is attributable to
the disruption of clouds across the L/T transition.  Their complex behavior
limits the use of the FeH(B) and CrH(B) indices to spectral
types T$_o$5 and later.
Finally, the Color-e index measures the red spectral
slope due to K I absorption, but is defined at longer wavelengths than the Color-d
\citep{kir99} or PC3 \citep{mrt96} indices because of the absence of
detectable flux shortward of 8000 {\AA}
in many of the spectra.  This index is useful earlier than roughly T$_o$2 (i.e., around the
L/T transition), but appears to
saturate (with considerable scatter) in the mid- and late-type T dwarfs,
possibly due to a turnover in the strength of the K I red wing (however, see $\S$5.2).
Taking into account similar behavior between pairs of indices
and differing sensitivity for early and late T types, we chose to use the CsI(A), FeH(B),
Color-e, and hybrid CrH(A)/H$_2$O ratios for our classification.
These ratio values are plotted versus optical type in Figure 7.

Ratio values for the five spectral standards are listed in Table 6, and
values for the remaining T dwarf spectra in our sample, as well as the associated
standard types for each ratio, are listed in Table 7.
Final classifications were derived by averaging those ratio
types (FeH(B) was not used for types earlier than T$_o$5, and Color-e was not used for types
later than T$_o$2) and rounding off to the nearest whole subclass.
Individual ratio types for each object agree within roughly one standard class.
Note that SDSS 0837$-$0000
and SDSS 1021$-$0304 are assigned uncertain classifications because
of the substantial gaps between the nearest standard subclasses.
With the exception of these two objects, all of our
spectral ratio-derived classifications agree with
our visually-derived classifications within one subclass, implying that our suite of indices
accurately represent the overall spectral morphology.  Furthermore,
there is excellent agreement between the optical ratio classifications
and the near-infrared classifications, again with the singular exception of SDSS 0423$-$0414.

Based on these results, we conclude that classification of mid- and late-type
T dwarfs in the optical is feasible, either by visual comparison to the selected standards or
through the use of spectral indices defined in Table 5.
However, our scheme is only accurate to within one subclass, and is therefore
not as precise as current near-infrared schemes.
The behavior of the Color-e and CrH(A)/H$_2$O
indices also suggest that early-type T dwarfs may be delineated in the optical,
suggesting that a clear segregation of L and T dwarfs
at these wavelengths is possible.  Optical spectral observations of
additional objects in this
spectral type range will have to be made, however.  We note that
disagreement between the optical and near-infrared
spectral morphologies of the key L/T transition object,
SDSS 0423$-$0414, does point to possible difficulties in this regime \citep{kir03}.

\section{Discussion}

\subsection{2MASS 0937+2931: A Metal-Poor Halo/Thick Disk Brown Dwarf?}

The only object in our sample whose red optical spectrum is truly unique
in comparison to the
spectral standards is the T$_n$6p/T$_o$7p 2MASS 0937+2931.
This object stands out in the near-infrared because of
its enhanced CIA H$_2$ absorption (resulting in a highly suppressed K-band peak)
and absence of the 1.25 $\micron$ K I lines, both of which point to
a metal-poor and/or high gravity atmosphere \citep{me02a}.
As noted in $\S$ 4.2, the optical spectrum is
peculiar because of its strong spectral slope, also a consequence
of low atmospheric metallicity and/or high surface gravity \citep{bur02}.

To examine the relative contribution of these secondary effects,
we have compared the spectrum of 2MASS 0937+2931
to theoretical models from \citet{bur02}.
These models do not incorporate the most recent
line broadening theory of \citet{bur03}, nor do they include FeH absorption,
a relatively strong feature in the spectrum of 2MASS 0937+2931.
Nevertheless, as the current state of the art, they serve to illustrate general trends.
As discussed in \citet{bur02}, increased surface gravity and decreased metallicity
both result in a higher pressure photosphere, and hence enhancement of
the pressure-sensitive K I and Na I line wings.  This effect is strong enough
to overcome the reduced column depth and decreased chemical abundance resulting
from a higher gravity or lower metallicity atmosphere, respectively, which tend
to weaken other chemical features,
such as the 9250 {\AA} H$_2$O band.  Decreased temperature also strengthens the
Na I and K I lines, but results in stronger H$_2$O bands as well (see $\S$ 5.2).
When we compare these trends to the optical spectrum of 2MASS 0937+2931
(Figure 8), it becomes clear that the relatively weak H$_2$O band seen in
these data requires a moderately high T$_{eff}$, while the strong Na I and K I
require both a high gravity and low metallicity atmosphere.
The best fitting models from \citet{bur02},
T$_{eff}$ = 800 and 1000 K, g = $3{\times}10^5$ cm s$^{-2}$, and $Z = 0.3 Z_{\sun}$
are shown in Figure 8 along with the observed spectrum.
Both models show somewhat stronger inner K I wings than the data, consistent
with the analysis of \citet{bur03}
that the line-broadening theory used in \citet{bur02} overestimates
the line opacity by as much as a factor of 10 shortward of 9000 {\AA}.
Nevertheless, the reasonable fit of the models to the data
indicate a low metallicity, high
gravity atmosphere for this object.
Note that the constraint on T$_{eff}$ is fairly weak;
a robust estimate could be obtained by measurement of this object's parallax.

The substantial
proper motion of 2MASS 0937+2931 ($\mu = 1{\farcs}4{\pm}0{\farcs}2$;
Burgasser et al.\ 2003a) suggests that it is either quite close to the Sun or
has a high space velocity.  \citet{me03a} have estimated a
J-band spectrophotometric distance
of roughly 9 pc, implying $v_{tan}$ $\approx$ 60 km s$^{-1}$,
which suggests (but does not guarantee) membership in the
Galactic thick disk or halo populations.
Such an object would have both a high surface gravity (as it must be both old
and therefore more massive for its temperature)
and a metal-deficient atmosphere \citep{giz97}.
\citet{me03c} have recently identified an L-type metal-deficient dwarf which, like
2MASS 0937+2931, exhibits strong K I and metal hydride absorption, relatively
blue near-infrared colors, and substantial space motion.  Determining whether
2MASS 0937+2931 is a cool analog to this substellar subdwarf requires additional
astrometric data and refined spectral models.  Nonetheless,
the observed peculiarities in its optical and near-infrared spectra
clearly indicate that 2MASS 0937+2931 is a unique T dwarf requiring further investigation.

\subsection{Disentangling Gravity and Temperature Effects in T Dwarf Optical Spectra}

If we ignore metallicity variations
(i.e., excluding 2MASS 0937+2931), then differences amongst the
spectra in our sample are determined largely by temperature and gravity effects.
The Na I and K I resonant doublet lines, which generally dominate this
spectral region, are highly sensitive to both parameters.  So too is the
9250 {\AA} H$_2$O band.  However, while the alkali lines strengthen with
increasing gravity, H$_2$O absorption decreases because of the reduced column
depth of the atmosphere \citep{bur02}.  The contrasting behavior of these features
suggests that they may be used to distinguish temperature and gravity effects
in T dwarf spectra.
As our sample likely includes objects with
ages of roughly 1--5 Gyr (assuming they are drawn from
the Galactic Disk population;
Reid \& Hawley 2000) and
700 $\lesssim$ T$_{eff}$ $\lesssim$ 1200 K \citep{me02a},
and hence surface gravities ($g$) spanning the range
$(0.5-2.5){\times}10^5$ cm s$^{-2}$ \citep{bur97},
demonstrative gravity effects may be present.

As noted in $\S$ 4.3, the Color-e index (which measure the K I red wing)
shows substantial scatter with optical spectral type for classes later than T$_o$2
(Figure 7d), while the combined CrH(A)/H$_2$O index is reasonably monotonic
over this region.
Figure 9 compares the Color-e and CrH(A)/H$_2$O indices for all of our
telluric-calibrated spectra.
The general trends with T$_{eff}$ and gravity are noted at the bottom,
derived by measuring the indices on the spectral models of \citet{bur02}.
For CrH(A)/H$_2$O $\lesssim$ 0.7 (SpT $\gtrsim$ T$_o$2), there is a clear
spread in the Color-e values, with 2MASS 0937+2931 being the most
significant outlier.  The simplest explanation for this divergence is intrinsic scatter in
the indices, due to perhaps poor signal-to-noise in some of the spectra.
However, we note two suggestive trends.  First, the four
objects with H$\alpha$ emission (including SDSS 0423$-$0414; Kirkpatrick et al.\ 2003)
generally have steeper spectral slopes (larger Color-e) than other
objects with similar CrH(A)/H$_2$O values.  \citet{giz00} have noted that for late-type M and
L field dwarfs, H$\alpha$ emission is generally found amongst the older,
and hence more massive,
objects. If this relation follows through into the T dwarf regime, then we expect the
active objects to have higher surface gravities, and therefore systematically larger Color-e ratios.
Second, one object in our sample has an independent age, and hence gravity,
determination, the companion T$_n$8/T$_o$7 Gliese 570D \citep{me00a}.
At 2--5 Gyr, this brown dwarf has a surface gravity of (1--2)${\times}10^5$ cm s$^{-2}$
\citep{geb01}.
Below this object is the T$_n$7.5/T$_o$7 2MASS 1217$-$0311 \citep{me99}, which
\citet{leg03} have suggested may be a low-gravity T dwarf based on its
somewhat brighter K-band peak (implying weaker CIA H$_2$ absorption).
Again, the relative positions of these sources in Figure 9 are consistent with
the expected trends.

These examples suggest that temperature- and gravity-based features could be
disentangled, enabling a means of resolving the brown dwarf age/mass/temperature degeneracy.
Independent age and/or gravity information (e.g., kinematics,
companionship to well-studied stars, or cluster membership)
or improved spectral models are needed to verify and
calibrate these potentially useful diagnostics.

\subsection{Comparison of Optical and Near-Infrared Classifications}

One of our motivations for deriving a classification scheme for T dwarfs in
the optical is to compare the evolution of spectral morphology between optical
and near-infrared wavelengths.  For this discussion, we focus on the
mid- and late-type T dwarfs, T$_n$5/T$_o$5 and later.
As discussed above, classifications for these objects in the two wavelength regimes
considered here are
remarkably consistent, implying a coherent evolution of features over
0.6--2.5 $\micron$ with spectral type.
This may be somewhat surprising,
given the competing temperature and gravity dependencies of
defining H$_2$O and CH$_4$ bands \citep{bur99} and K I and Na I
resonance line wings \citep{bur02},
the relatively weak gravity dependence of the optical
Cs I lines \citep{lod99}, and the strong gravity dependence of CIA H$_2$ in
the near-infrared \citep{bor97}.
It may be the case that, despite the wide range of possible surface gravities
in our sample, these particular T dwarfs have fairly similar masses and ages
and hence little variation in $g$.  Such a case would imply that
temperature is the $de facto$
discriminant in current T dwarf classification schemes.
However, the suggestive trends noted in the previous section do point to
separable gravity/temperature diagnostics being present within our sample,
and hence a non-negligible range of surface gravities.  This possibility
leads us to consider
that competing gravity and temperature effects may
compensate for each other amongst all of the optical and near-infrared
diagnostic spectral features, and that
T subtypes in either regime may not in fact map directly onto a temperature scale.
The latter point is an
important problem, as classifications are
commonly used as proxies for T$_{eff}$ and luminosity in population studies.
Our results indicate that comparing optical and near-infrared morphologies
cannot, at the precision of our near-infrared and optical
classifications, resolve dependencies between gravity and temperature
in the spectra of mid- and late-type T dwarfs.
However, developing the spectral diagnostics described in previous section
may lead to a two-dimensional system that could be
tied directly to temperature and gravity scales.

The agreement between optical and near-infrared classifications
for mid- and late-type T dwarfs also suggests that two possible causes
for disagreements amongst late-type L and early-type T dwarf classifications --
condensate cloud opacity and/or duplicity \citep{kir03} -- may not be important
for later-type T dwarfs.
In general, T dwarf spectra and photometry are well-matched to models without
condensate cloud opacity
\citep[however, see Marley et al.\ 2002; Burrows et al.\ 2002a]{tsu96,tsu99,all96,all01,bur97,ack01,tsu02},
although low-level near-infrared photometric \citep{eno03} and spectroscopic \citep{nak00}
variability, attributed to condensate clouds, has been observed in some T dwarfs.
These time-dependent variations are typically quite small, however, and
do not significantly change the overall spectral morphology \citep{nak00}.
As for duplicity, later-type T dwarf doubles
are not affected by the substantial redistribution of spectral flux that occurs
across the L/T transition \citep{dah02}, which allows a late-type L dwarf to dominate
optical emission while its early-type T dwarf companion dominates near-infrared
emission.  This scenario may explain the presence of CH$_4$ at 2.2 $\micron$
in the L6.5 unequal-magnitude binary 2MASS 0920+3517AB \citep{kir00,rei01,nak01},
as well as the near-infrared/optical
classification discrepancy for SDSS 0423$-$0414 \citep{kir03}.
In contrast, both of the known binaries in our sample, 2MASS 1225$-$2739AB and 2MASS 1534$-$2952AB
\citep{me03b}, have essentially identical optical and near-infrared types.
It is important to consider, however, that a larger, and perhaps time-resolved,
optical sample is required to rigorously
examine the influence of variability and duplicity on differences between optical and near-infrared
T dwarf spectra.

\section{Summary}

We have examined LRIS red optical spectra for a sample of T dwarfs.
These spectra are largely dominated by pressure-broadened
Na I and K I features, which are highly sensitive to temperature, gravity, and metallicity.
Weaker lines of Cs I and Rb I are also present, and appear to
follow the trends of current chemical equilibrium calculations. Molecular bands of
H$_2$O, CaH, CrH, FeH, and possibly
CH$_4$ are also present.  H$\alpha$ emission is seen in three objects, weakly in
SDSS 1254$-$0122 and 2MASS 1047+2124, but strong in the active T dwarf 2MASS 1237+6526.
None of the spectra exhibit the 6708 {\AA} Li I line, consistent with chemical depletion of
this atomic species for T $\lesssim$ 1500 K.
Trends in the observed strengths of these features have allowed
us to define a classification scheme in the red optical, accurate to within one
subclass.  Tying our scheme to selected near-infrared spectral standards, we are able to
make a consistent comparison between optical and near-infrared types, and hence
a critical examination of spectral morphologies in these two spectral regions.
We find excellent agreement between these classifications amongst the mid- and late-type
T dwarfs, suggesting
that either current classification schemes are largely temperature-based, or that
temperature and gravity effects compensate for each other throughout the 0.6--2.5 $\micron$ region.
Disentangling these physical parameters may be possible by comparing the strength
of the K I red wing and the 9250 {\AA} H$_2$O band, potentially leading to a two-dimensional
classification system that can be mapped onto temperature and gravity scales.
Both gravity and metallicity effects help explain the peculiar spectrum of
2MASS 0937+2931, which may be an old, massive, and slightly metal-poor thick disk or halo
brown dwarf.  Further astrometric data and improved spectral models will enable a more
rigorous characterization of
this unique brown dwarf.
Clearly, while very little of the total luminosity of T dwarfs is
emitted shortward of 1 $\micron$, there are an abundance of diagnostics
present in this spectral region that can be used
to examine the physics of these cool brown dwarfs.

\acknowledgments

We are grateful to our Keck Observing Assistants Joel Aycock, Julie Rivera,
and Terry Stickel, and Instrument Specialists/Support Astronomers Paola Amico,
Tom Bida, and Bob Goodrich for their
assistance during the data acquisition; and to the NASA
TAC for its allocation of time for this project.
We also thank our anonymous referee for her/his insightful comments and
criticisms.
A.\ J.\ B.\ acknowledges support provided by NASA through
Hubble Fellowship grant HST-HF-01137.01 awarded by the Space Telescope Science Institute,
which is operated by the Association of Universities for Research in Astronomy,
Incorporated, under NASA contract NAS5-26555.
J.\ D.\ K. acknowledges the support
of the Jet Propulsion
Laboratory, California Institute of Technology, which is operated under
contract with the National Aeronautics and Space Administration.
A.\ B.\ acknowledges funding through NASA grants NAG5-10760 and NAG5-10629.
This publication makes use of data from the Two
Micron All Sky Survey, which is a joint project of the University
of Massachusetts and the Infrared Processing and Analysis Center,
funded by the National Aeronautics and Space Administration and
the National Science Foundation.
2MASS data were obtain through
the NASA/IPAC Infrared Science Archive, which is operated by the
Jet Propulsion Laboratory, California Institute of Technology,
under contract with the National Aeronautics and Space
Administration.
Portions of the data presented herein were obtained at the
W.\ M.\ Keck Observatory which is operated as a scientific partnership
among the California Institute of Technology, the University of California,
and the National Aeronautics and Space Administration. The Observatory was
made possible by the generous financial support of the W.\ M.\ Keck Foundation.
The authors wish to recognize and acknowledge the very significant cultural role
and reverence that the summit of Mauna Kea has always had within the indigenous
Hawaiian community.  We are most fortunate to have the opportunity to conduct
observations from this mountain.

\clearpage

\begin{deluxetable}{lllccllc}
\tabletypesize{\scriptsize}
\tablecaption{Log of LRIS Observations.}
\tablewidth{0pt}
\tablehead{
\colhead{} &
\colhead{SpT} &
\colhead{} &
\colhead{} &
\colhead{} &
\colhead{} \\
\colhead{Object\tablenotemark{a}} &
\colhead{NIR/Opt.\tablenotemark{b}} &
\colhead{UT Date} &
\colhead{t$_{int}$ (s)} &
\colhead{Airmass} &
\colhead{Telluric Cal.} &
\colhead{Type\tablenotemark{c}} &
\colhead{Ref} \\
\colhead{(1)} &
\colhead{(2)} &
\colhead{(3)} &
\colhead{(4)} &
\colhead{(5)} &
\colhead{(6)} &
\colhead{(7)} &
\colhead{(8)} \\
}
\startdata
2MASS J04151954$-$0935066 & T$_n$8/T$_o$8 & 2001 Feb 20 & 6000 & 1.20--1.38  & WD 0413-074 & DA4 & 1 \\
2MASS J05591914$-$1404488 & T$_n$5/T$_o$5 & 2000 Mar 5 & 3600 & 1.20--1.22 & WD 0552-041 & DC9 & 2 \\
2MASS J07271824+1710012 & T$_n$7/T$_o$8 & 2000 Mar 5 & 1800 & 1.00 &  WD 0747+073.1 & DC9 & 1 \\
2MASS J07554795+2212169 & T$_n$5:/T$_o$6 & 2001 Feb 20 & 2400 & 1.13--1.19 &  SAO 79820 & G0 V & 1 \\
2MASS J09373487+2931409 & T$_n$6p/T$_o$7 & 2000 Mar 5 & 3600 & 1.01 &  (WD 0924+199)\tablenotemark{d} & (DC5)\tablenotemark{d} & 1 \\
2MASS J10475385+2124234 & T$_n$6.5/T$_o$7 & 2001 Feb 20 & 4800 & 1.06--1.22  & HD 93583 & G0 V & 3 \\
2MASS J12171110$-$0311131 & T$_n$7.5/T$_o$7 & 2000 Mar 5 & 1800 & 1.11--1.16  & WD 1225-079 & DZA5 & 3 \\
2MASS J12255432$-$2739466AB & T$_n$6/T$_o$6 & 2000 Mar 5 & 3600 & 1.48--1.50 & WD 1225-079 & DZA5 & 3 \\
2MASS J12373919+6526148 & T$_n$6.5/T$_o$7 & 2000 Mar 5 & 1800 & 1.46 &  & & 3 \\
 & & 2001 Feb 20 & 2400 & 1.49--1.52 &  SAO 15828 & F8 V & 3 \\
SDSSp J125453.90$-$012247.4 & T$_n$2/T$_o$2 & 2001 Feb 20 & 3600 & 1.08--1.13 & HD 111942 & G0 V & 4 \\
Gliese 570D & T$_n$8/T$_o$7 & 2000 Mar 5 & 3600 & 1.33--1.36 &  WD 1444-174 & DC9 & 5 \\
 & & 2001 Feb 20 & 4800 & 1.08--1.13 & HD 131878 & G0 V & 5 \\
2MASS J15031961+2525196 & T$_n$5.5/T$_o$6 & 2003 Jan 2 & 1200 & 1.35--1.40 & BD+26 2652 & G0 & 6 \\
2MASS J15344984$-$2952274AB & T$_n$5.5/T$_o$6 & 2001 Feb 20 & 2400 & 1.58--1.63 & HD 138874 & F7 V & 1 \\
\enddata
\tablenotetext{a}{2MASS Point Source Catalog source designations
are given as ``2MASS Jhhmmss[.]ss$\pm$ddmmss[.]s''. The
suffix conforms to IAU nomenclature convention and is the
sexagesimal Right Ascension and declination at J2000 equinox.}
\tablenotetext{b}{Near-infrared spectral types from \citet{me02a}; optical
spectral types from Tables 6 and 7.}
\tablenotetext{c}{White dwarf spectral types from \citet{mco99} and references therein.}
\tablenotetext{d}{A background F/G dwarf background
may have actually been observed; see $\S$ 2.}
\tablerefs{
(1) \citet{me02a}; (2) \citet{me00c}; (3) \citet{me99};
(4) \citet{leg00}; (5) \citet{me00a}; (6) \citet{me03a}.}
\end{deluxetable}

\begin{deluxetable}{lccl}
\tabletypesize{\scriptsize}
\tablecaption{Red Optical Features in T Dwarfs.}
\tablewidth{0pt}
\tablehead{
\colhead{Feature} &
\colhead{$\lambda$ ($\micron$)} &
\colhead{Transition}  &
\colhead{Ref.} \\
\colhead{(1)} &
\colhead{(2)} &
\colhead{(3)}  &
\colhead{(4)} \\
}
\startdata
Na I & 5890\tablenotemark{a} & 3s $^2$S$_{1/2}$ $-$ 3p $^2$P$_{3/2}$  & 1,2 \\
Na I & 5896\tablenotemark{a} & 3s $^2$S$_{1/2}$ $-$ 3p $^2$P$_{1/2}$  & 1,2 \\
H$\alpha$ & 6563 & 3d $^2$D$_{5/2}$ $-$ 2p $^2$P$_{3/2}$  & 1 \\
CaH & 6750--7050 & 0-0 band of A$^2$$\Pi$-X$^2$$\Sigma$ & 3 \\
K I & 7665\tablenotemark{a} & 4s $^2$S$_{1/2}$ $-$ 4p $^2$P$_{3/2}$  & 1,2 \\
K I & 7699\tablenotemark{a} & 4s $^2$S$_{1/2}$ $-$ 4p $^2$P$_{1/2}$  & 1,2 \\
Rb I & 7800 & 5s $^2$S$_{1/2}$ $-$ 5p $^2$P$_{3/2,1/2}$  & 1 \\
Rb I & 7948 & 5s $^2$S$_{1/2}$ $-$ 5p $^2$P$_{3/2,1/2}$  & 1 \\
CH$_4$\tablenotemark{b} & 8800--9200 & 4($\nu$$_1$,$\nu$$_3$) & 4 \\
Cs I & 8521 & 6s $^2$S$_{1/2}$ $-$ 6p $^2$P$_{3/2}$ & 1 \\
CrH & 8611 bandhead & 0-0 band of A$^6$$\Sigma$$^+$-X$^6$$\Sigma$$^+$ & 5 \\
FeH & 8692 bandhead & 1-0 band of A$^4$$\Delta$-X$^4$$\Delta$ & 6 \\
Cs I & 8943 & 6s $^2$S$_{1/2}$ $-$ 6p $^2$P$_{1/2}$ & 1 \\
H$_2$O & 9250--9400 & 3($\nu$$_1$,$\nu$$_3$) & 7 \\
H$_2$O & 9450--9800 & 2($\nu$$_1$,$\nu$$_3$) + 2$\nu$$_2$ & 7 \\
FeH & 9896 bandhead & 0-0 band of A$^4$$\Delta$-X$^4$$\Delta$ & 6 \\
CrH & 9969 bandhead & 0-1 band of A$^6$$\Sigma$$^+$-X$^6$$\Sigma$$^+$ & 5  \\
\enddata
\tablenotetext{a}{Pressure-broadened over $>$ 1000 {\AA}.}
\tablenotetext{b}{Detection of this feature is ambiguous; see $\S$3.2 and Figure 4.}
\tablerefs{
(1) \citet{wie66}; (2) \citet{bur00}; (3) \citet{ber74};
(4) \citet{dic77}; (5) \citet{kle59} (6) \citet{win77}; (7) \citet{aum67}.}
\end{deluxetable}

\begin{deluxetable}{llccccccccccc}
\tabletypesize{\scriptsize}
\rotate
\tablecaption{Cs I and Rb I Pseudo-Equivalent Widths.}
\tablewidth{0pt}
\tablehead{
 & \colhead{SpT} & \multicolumn{2}{c}{8521 {\AA} Cs I} &
& \multicolumn{2}{c}{8943 {\AA} Cs I} &
& \multicolumn{2}{c}{7800 {\AA} Rb I} &
& \multicolumn{2}{c}{7948 {\AA} Rb I} \\
\cline{3-4} \cline{6-7} \cline{9-10} \cline{12-13}
\colhead{Object} &
\colhead{NIR/Opt\tablenotemark{a}} &
\colhead{${\lambda}_c$ ({\AA})} &
\colhead{$pEW$ ({\AA})} & &
\colhead{${\lambda}_c$ ({\AA})} &
\colhead{$pEW$ ({\AA})} & &
\colhead{${\lambda}_c$ ({\AA})} &
\colhead{$pEW$ ({\AA})} & &
\colhead{${\lambda}_c$ ({\AA})} &
\colhead{$pEW$ ({\AA})} \\
\colhead{(1)} &
\colhead{(2)} &
\colhead{(3)} &
\colhead{(4)} & &
\colhead{(5)} &
\colhead{(6)} & &
\colhead{(7)} &
\colhead{(8)} & &
\colhead{(9)} &
\colhead{(10)} \\
}
\startdata
SDSS 1254$-$0122 & T$_n$2/T$_o$2 & 8520$\pm$2 & 8.5$\pm$0.4  & & 8942$\pm$2 & 8.0$\pm$0.5 & & ... & $<$ 7 & &  7947$\pm$8 & 6.2$\pm$1.3 \\
2MASS 0559$-$1404 & T$_n$5/T$_o$5 & 8522$\pm$2 & 8.1$\pm$0.5  & & 8945$\pm$2 & 6.9$\pm$0.3 & & 7800$\pm$30 & 6$\pm$5 & & 7950$\pm$14 & 11.6$\pm$1.8 \\
2MASS 0755+2212 & T$_n$5:/T$_o$6 & 8521$\pm$3 & 6.2$\pm$0.6  &  & 8944$\pm$2 & 7.5$\pm$0.5 & & ... & $<$ 2 & & 7949$\pm$11 & 4.3$\pm$2.1 \\
2MASS 1503+2525 & T$_n$5.5/T$_o$6 & 8519$\pm$2 & 7.1$\pm$0.3  & & 8941$\pm$2 & 8.0$\pm$0.3 & & 7800$\pm$30 & 12$\pm$5 & & 7944$\pm$15 & 7.1$\pm$1.6 \\
2MASS 1534$-$2952AB & T$_n$5.5/T$_o$6 & 8517$\pm$2 & 8.7$\pm$1.2  & & 8940$\pm$2 & 7.7$\pm$0.6 & & ... & $<$ 130 & & 7943$\pm$8 & 15$\pm$12 \\
2MASS 1225$-$2739AB & T$_n$6/T$_o$6 & 8522$\pm$2 & 9.1$\pm$1.3  & & 8944$\pm$2 & 8.5$\pm$0.8 & & ... & $<$ 24 & & ... & $<$ 10 \\
2MASS 0937+2931 & T$_n$6p/T$_o$7 p & 8520$\pm$2 & 4.8$\pm$0.8  & & 8942$\pm$2 & 7.7$\pm$0.5 & & ... & ... & & 7949$\pm$17 & 10$\pm$6 \\
2MASS 1047+2124 &  T$_n$6.5/T$_o$7 &  8522$\pm$3 & 4.9$\pm$0.9 &  & 8946$\pm$4 & 5.7$\pm$0.9 & & ... & $<$ 25 & & ... & $<$ 7 \\
2MASS 1237+6526\tablenotemark{b} & T$_n$6.5/T$_o$7 & ... & $<$ 9  &  & 8941$\pm$3 & 8$\pm$4 & & ... & ... & & ... & $<$ 60 \\
 &  & 8522$\pm$3 & 6.4$\pm$1.2  & & 8943$\pm$2 & 6.0$\pm$0.7 & & ... & $<$ 40 & & ... & $<$ 14 \\
2MASS 0727+1710 & T$_n$7/T$_o$8 & 8519$\pm$3 & 4.7$\pm$0.9  & & 8943$\pm$2 & 7.3$\pm$0.7 & & ... & $<$ 50 & & 7949$\pm$10 & 16$\pm$7 \\
2MASS 1217$-$0311 & T$_n$7.5/T$_o$7 & 8521$\pm$2 & 5.6$\pm$1.4  &  & 8944$\pm$3 & 4.0$\pm$0.6 & & ... & $<$ 18 & & ... & $<$ 8 \\
Gliese 570D\tablenotemark{b} & T$_n$8/T$_o$7 & 8519$\pm$2 & 4.2$\pm$1.0  & & 8942$\pm$2 & 5.3$\pm$0.9 & & ... & $<$ 18 & & ... & ... \\
 &  & 8523$\pm$3 & 4.0$\pm$0.6  & & 8944$\pm$2 & 6.7$\pm$0.9 & & ... & $<$ 18 & & 7951$\pm$7 & 11$\pm$7  \\
2MASS 0415$-$0935 & T$_n$8/T$_o$8 & 8520$\pm$9 & 1.8$\pm$1.4  & & 8943$\pm$2 & 3.7$\pm$1.1 & & ... & $<$ 12 & & 7948$\pm$7 & 11$\pm$6 \\
\enddata
\tablenotetext{a}{Near-infrared spectral types from \citet{me02a}; optical spectral
types from Tables 6 and 7.}
\tablenotetext{b}{First set of values from epoch 2000 March 5 (UT), second set of
values from epoch 2001 February 20 (UT).}
\end{deluxetable}

\begin{deluxetable}{llcc}
\tabletypesize{\scriptsize}
\tablecaption{H$\alpha$ Emission Strengths.}
\tablewidth{0pt}
\tablehead{
\colhead{} &
\colhead{SpT} &
\colhead{$f_{H{\alpha}}$} &
\colhead{} \\
\colhead{Object} &
\colhead{NIR/Opt\tablenotemark{a}} &
\colhead{(10$^{-18}$ ergs cm$^{-2}$ s$^{-1}$)} &
\colhead{$\log{L_{H{\alpha}}/L_{bol}}$} \\
\colhead{(1)} &
\colhead{(2)} &
\colhead{(3)} &
\colhead{(4)} \\
}
\startdata
SDSS 1254$-$0122 & T$_n$2/T$_o$2 & 7.5$\pm$2.5  & $-5.8$  \\
2MASS 0559$-$1404 & T$_n$5/T$_o$5 & $< 6.1$  & $< -6.1$  \\
2MASS 0755+2212 & T$_n$5:/T$_o$6 & $< 12$  & $< -5.1$  \\
2MASS 1503+2525 & T$_n$5.5/T$_o$6 & $< 9.6$  & $< -5.5$  \\
2MASS 1534$-$2952AB & T$_n$5.5/T$_o$6 & $< 17$  & $< -5.2$  \\
2MASS 1225$-$2739AB & T$_n$6/T$_o$6 & $< 6.7$  & $< -5.4$  \\
2MASS 0937+2931 & T$_n$6p/T$_o$7 p & $< 3.9$  & $< -6.0$  \\
2MASS 1047+2124 &  T$_n$6.5/T$_o$7 & 5.9$\pm$2.7  & $-5.4$  \\
2MASS 1237+6526\tablenotemark{b} & T$_n$6.5/T$_o$7 & 29.0$\pm$2.8\tablenotemark{c}  & $-4.6$  \\
 &  & 105.8$\pm$2.8\tablenotemark{c}  & $-4.1$  \\
2MASS 0727+1710 & T$_n$7/T$_o$8 & $< 3.6$  & $< -5.7$  \\
2MASS 1217$-$0311 & T$_n$7.5/T$_o$7 & $< 7.7$  & $< -5.3$  \\
Gliese 570D\tablenotemark{b} & T$_n$8/T$_o$7 & $< 6.5$  & $< -5.3$  \\
 &  & $< 9.0$  & $< -5.1$   \\
2MASS 0415$-$0935 & T$_n$8/T$_o$8 & $<$ 7.9  & $< -5.4$  \\
\enddata
\tablenotetext{a}{Near-infrared spectral types from \citet{me02a}; optical spectral
types from Tables 6 and 7.}
\tablenotetext{b}{First set of values from epoch 2000 March 5 (UT), second set of
values from epoch 2001 February 20 (UT).}
\tablenotetext{c}{See \citet{me02b} for a discussion on the variability of the H$\alpha$ line in
2MASS 1237+6526.}
\end{deluxetable}

\begin{deluxetable}{lcccl}
\tabletypesize{\scriptsize}
\tablecaption{ T Dwarf Spectral Ratios in the Red Optical.}
\tablewidth{0pt}
\tablehead{
\colhead{Diagnostic} &
\colhead{Numerator ({\AA})\tablenotemark{a}} &
\colhead{Denominator ({\AA})\tablenotemark{a}} &
\colhead{Feature Measured} &
\colhead{Ref.} \\
\colhead{(1)} &
\colhead{(2)} &
\colhead{(3)} &
\colhead{(4)} &
\colhead{(5)} \\
}
\startdata
CsI(A) & ${\langle}F_{8496.1-8506.1}{\rangle}$+${\langle}F_{8536.1-8546.1}{\rangle}$ & $2{\times}{\langle}F_{8516.1-8526.1}{\rangle}$ & 8521 {\AA} Cs I & 1,2\tablenotemark{b} \\
CsI(B) & ${\langle}F_{8918.5-8928.5}{\rangle}$+${\langle}F_{8958.3-8968.3}{\rangle}$ & $2{\times}{\langle}F_{8938.5-8948.3}{\rangle}$ & 8943 {\AA} Cs I & 1,2\tablenotemark{b} \\
H$_2$O & $\int{F_{9220-9240}}$ & $\int{F_{9280-9300}}$ & 9250 {\AA} H$_2$O  & 2 \\
CrH(A) & $\int{F_{8560-8600}}$ & $\int{F_{8610-8650}}$ & 8611 {\AA} CrH  & 3 \\
CrH(B) & $\int{F_{9855-9885}}$ & $\int{F_{9970-10000}}$ & 9969 {\AA} CrH  & 2,3\tablenotemark{c} \\
FeH(A) & $\int{F_{8560-8600}}$ & $\int{F_{8685-8725}}$ & 8692 {\AA} FeH  & 3 \\
FeH(B) & $\int{F_{9855-9885}}$ & $\int{F_{9905-9935}}$ & 9896 {\AA} FeH  & 2,3\tablenotemark{c} \\
Color-e & ${\langle}F_{9140-9240}{\rangle}$ & ${\langle}F_{8400-8500}{\rangle}$ & Spectral Slope & 2  \\
\enddata
\tablenotetext{a}{${\langle}{\rangle}$ denotes average of flux density in range specified;
$\int$ denotes integrated flux in range specified.}
\tablenotetext{b}{These indices are essentially identical to Cs-a and Cs-b defined by \citet{kir99}.}
\tablenotetext{c}{Modified from ratios defined by \citet{mrt99}.}
\tablerefs{(1) \citet{kir99}; (2) This paper; (3) \citet{mrt99}.}
\end{deluxetable}

\begin{deluxetable}{llccccl}
\tabletypesize{\scriptsize}
\tablecaption{Standard Spectral Ratio Values.}
\tablewidth{0pt}
\tablehead{
\colhead{Object} &
\colhead{Opt SpT} &
\colhead{CsI(A)} &
\colhead{CrH(A)/H$_2$O} &
\colhead{FeH(B)} &
\colhead{Color-e} &
\colhead{Ref} \\
\colhead{(1)} &
\colhead{(2)} &
\colhead{(3)} &
\colhead{(4)} &
\colhead{(5)} &
\colhead{(6)} &
\colhead{(7)} \\
}
\startdata
2MASS 1632+1904 & L8 & 1.70 & 1.02 & 1.11 & 1.88 & 1 \\
SDSS 1254$-$0122\tablenotemark{a} & T$_o$2 & 2.01 & 0.78 & 1.13 & 4.02 & 2,3 \\
2MASS 0559$-$1404 & T$_o$5 & 1.77 & 0.63 & 1.37 & 4.24 & 2 \\
SDSS 1624+0029 & T$_o$6 & 1.68 & 0.47 & 1.15 & 3.83 & 4 \\
2MASS 0415$-$0935\tablenotemark{a} & T$_o$8 & 1.19 & 0.25 & 0.94 & 4.20 & 2,3 \\
\enddata
\tablenotetext{a}{Measured from combined spectrum averaged over separate epochs.}
\tablerefs{(1) \citet{kir99}; (2) This paper; (3) \citet{kir03};
(4) \citet{me00b}.}
\end{deluxetable}

\begin{deluxetable}{lccccllll}
\rotate
\tabletypesize{\scriptsize}
\tablecaption{Late-type L and T Dwarf Spectral Ratio Values and Optical Spectral Types.}
\tablewidth{0pt}
\tablehead{
 & & & & &  \multicolumn{2}{c}{Optical SpT\tablenotemark{a}} & \colhead{} &  \\
\noalign{\smallskip} \cline{6-7}
\colhead{Object} &
\colhead{CsI(A)} &
\colhead{CrH(A)/H$_2$O} &
\colhead{FeH(B)} &
\colhead{Color-e} &
\colhead{Indices} &
\colhead{Visual} &
\colhead{NIR SpT\tablenotemark{b}} &
\colhead{Ref} \\
\colhead{(1)} &
\colhead{(2)} &
\colhead{(3)} &
\colhead{(4)} &
\colhead{(5)} &
\colhead{(6)} &
\colhead{(7)} &
\colhead{(8)} &
\colhead{(9)}  \\
}
\startdata
SD 0423$-$0414\tablenotemark{c} & 1.95 (T$_o$2) & 1.20 ($<$ L8) & 1.44 ($<$ T$_o$5) & 2.21 (L8)  & L8:\tablenotemark{d}  & $<$ L8\tablenotemark{d} & T$_n$0 & 1 \\
SD 0837$-$0000 & 1.80 (L8/T$_o$2) & 0.89 (L8/T$_o$2) & 0.90 ($<$ T$_o$5) & 3.35 (L8/T$_o$2) & T$_o$0${\pm}2$\tablenotemark{e}  & T$_o$2 & T$_n$1 & 1 \\
SD 1021$-$0304 & 1.98 (T$_o$2) & 0.67 (T$_o$2/5) & 1.37 (T$_o$5) & 3.85 ($\geq$ T$_o$2)  & T$_o$4${\pm}2$\tablenotemark{e}  & T$_o$2 & T$_n$3 &   1 \\
2M 1503+2525 & 1.61 (T$_o$6) & 0.50 (T$_o$6) & 1.30 (T$_o$5/6) & 4.75 ($\geq$ T$_o$2)  & T$_o$6  & T$_o$5/6 & T$_n$5.5 &  2 \\
2M 1534$-$2952AB & 1.70 (T$_o$6) & 0.54 (T$_o$5/6) & 1.38 (T$_o$5) & 4.47 ($\geq$ T$_o$2)  & T$_o$6  & T$_o$5/6 & T$_n$5.5 &  2 \\
2M 0755+2212 & 1.50 (T$_o$6/8) & 0.52 (T$_o$5/6) & 1.33 (T$_o$5) & 3.78 ($\geq$ T$_o$2)  & T$_o$6  & T$_o$5/6 & T$_n$5.5: &  2 \\
2M 1225$-$2739AB & 1.82 (T$_o$5) & 0.41 (T$_o$6) & 1.17 (T$_o$6) & 4.28 ($\geq$ T$_o$2)  & T$_o$6  & T$_o$6 & T$_n$6  & 2 \\
SD 1346$-$0031 & 1.43 (T$_o$6/8) & 0.35 (T$_o$6/8) & 1.01 (T$_o$6/8) & 5.01 ($\geq$ T$_o$2)  & T$_o$7  & T$_o$6/8  & T$_n$6  & 3 \\
2M 0937+2931 & 1.55 (T$_o$6/8) & 0.38 (T$_o$6/8) & 1.09 (T$_o$6/8) & 5.99 ($\geq$ T$_o$2)  & T$_o$7  & T$_o$6/8 p\tablenotemark{f} & T$_n$6 p & 2 \\
2M 1047+2124 & 1.46 (T$_o$6/8) & 0.37 (T$_o$6/8) & 1.00 (T$_o$6/8) & 4.41 ($\geq$ T$_o$2)  & T$_o$7  &  T$_o$6/8 & T$_n$6.5  & 2 \\
2M 1237+6526\tablenotemark{c} & 1.58 (T$_o$6) & 0.37 (T$_o$6/8) & 0.92 (T$_o$8) & 4.93 ($\geq$ T$_o$2)  & T$_o$7  & T$_o$6/8  & T$_n$6.5 & 2,3 \\
2M 0727+1710 & 1.28 (T$_o$8) & 0.36 (T$_o$6/8) & 0.92 (T$_o$8) & 3.84 ($\geq$ T$_o$2)  & T$_o$8  &  T$_o$6/8 & T$_n$7  & 2 \\
2M 1217$-$0311 & 1.71 (T$_o$5/6) & 0.34 (T$_o$6/8) & 0.92 (T$_o$8) & 3.77 ($\geq$ T$_o$2)  & T$_o$7  & T$_o$8 & T$_n$7.5  & 2 \\
Gliese 570D\tablenotemark{c} & 1.48 (T$_o$6/8) & 0.34 (T$_o$6/8) & 0.87 (T$_o$8) & 4.39 ($\geq$ T$_o$2)  & T$_o$7  & T$_o$8 & T$_n$8  &  1,2 \\
\enddata
\tablenotetext{a}{Optical spectral types quoted in the paper are those derived from the
spectral indices.}
\tablenotetext{b}{Near-infrared spectral types from \citet{me02a} except for SDSS 0423$-$0414
\citep{geb02}.}
\tablenotetext{c}{Measured from combined spectrum averaged over separate epochs.}
\tablenotetext{d}{Classified in the optical as L7.5 by \citet{kir03}.}
\tablenotetext{e}{These classifications are more uncertain due to the sparsity of standards between L8 and T5.}
\tablenotetext{f}{Slope of K I red wing significantly exceeds that of all T dwarf standards.}
\tablerefs{(1) \citet{kir03}; (2) This paper;
(3) \citet{me00b}.}
\end{deluxetable}

\clearpage

\begin{figure}
\plottwo{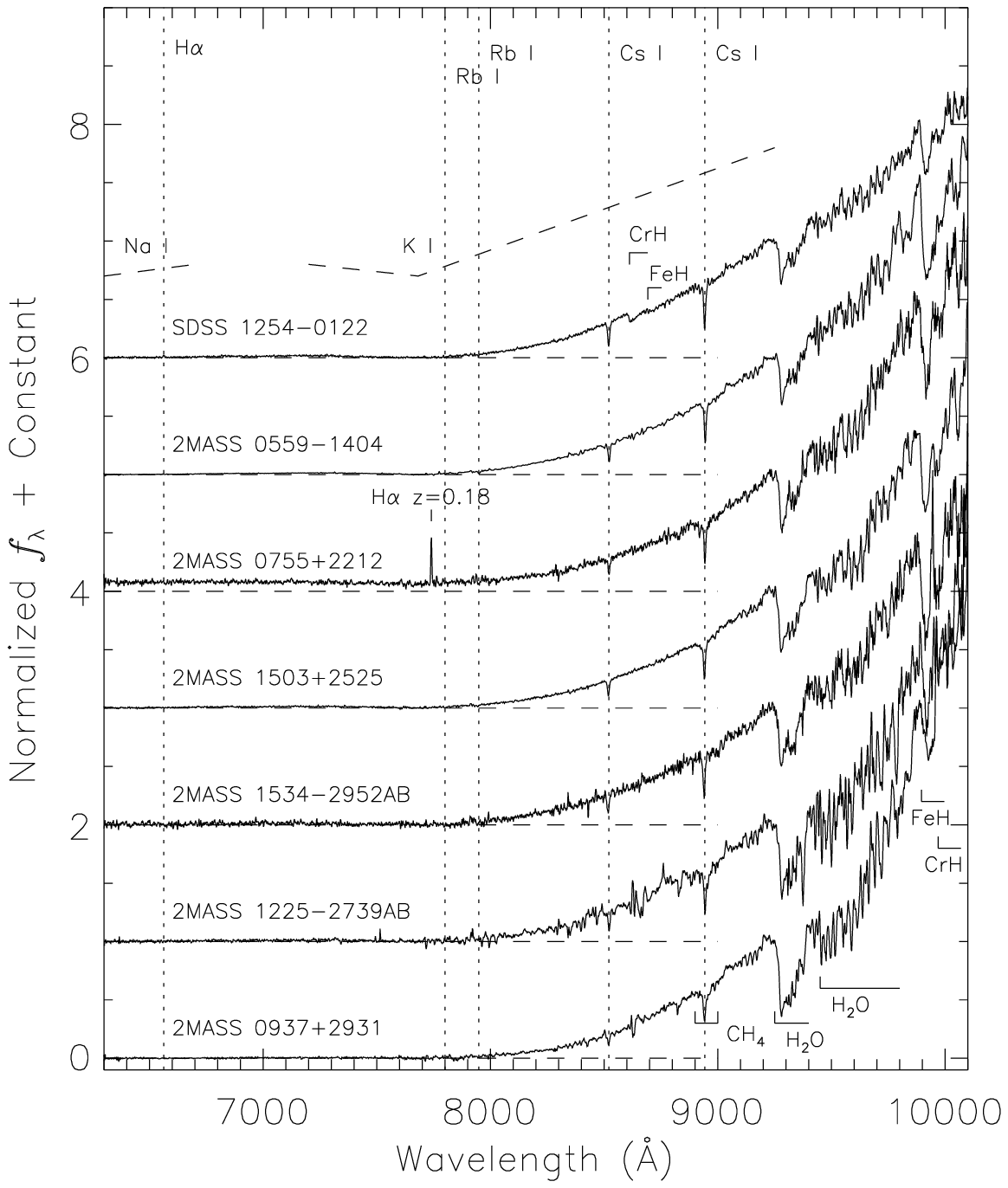}{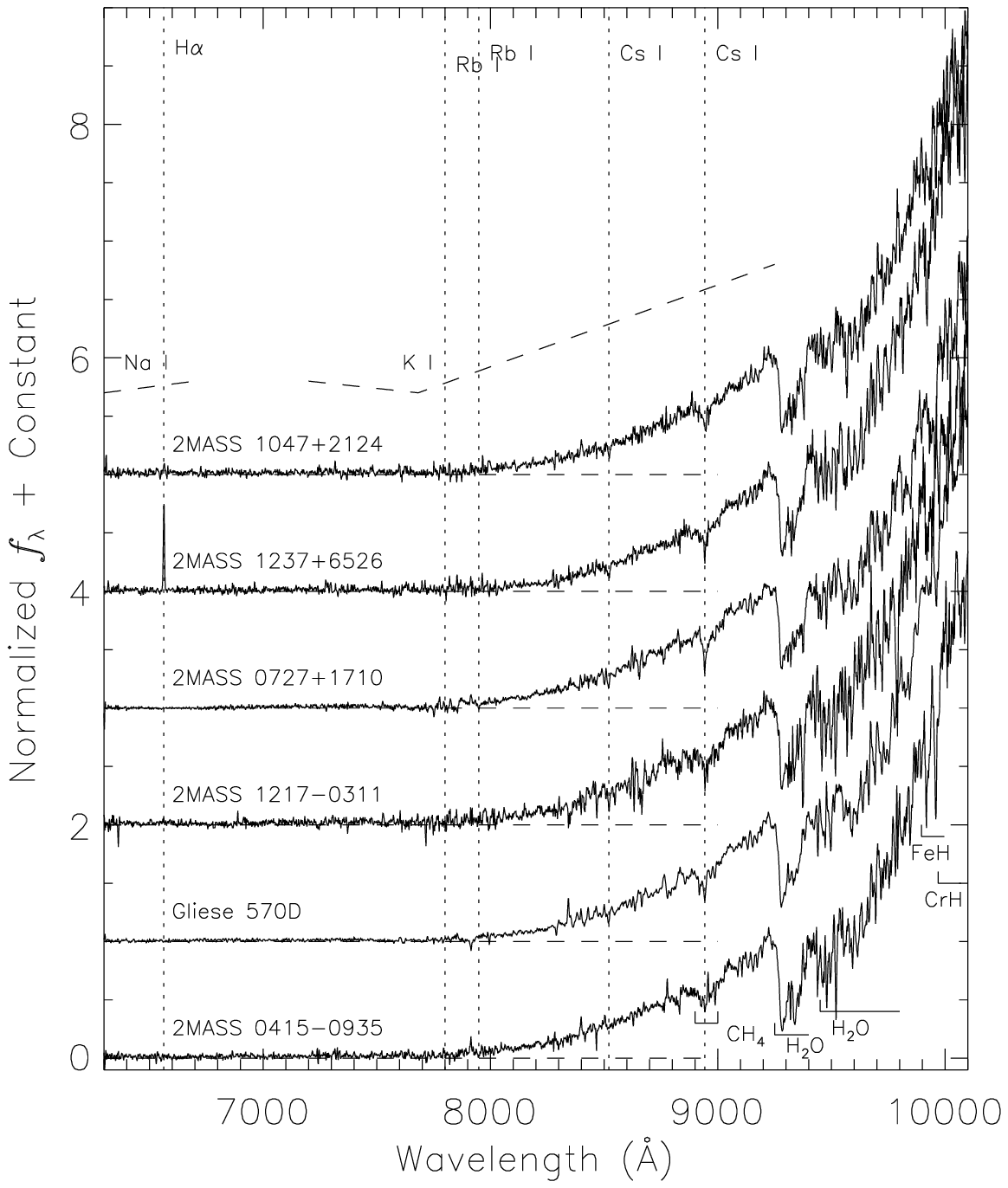}
\caption{Reduced red optical (6300--10100 {\AA}) spectra for
our T dwarf sample.
Data are normalized at 9250 {\AA} and offset by a constant (dashed lines).
Atomic and molecular absorption features listed in Table 2 are indicated.
Note that the emission feature in the spectrum of 2MASS 0755+2212 is
a redshifted ($z = 0.18$) H$\alpha$ line
from an aligned background galaxy.\label{fig1}}
\end{figure}

\begin{figure}
\plotone{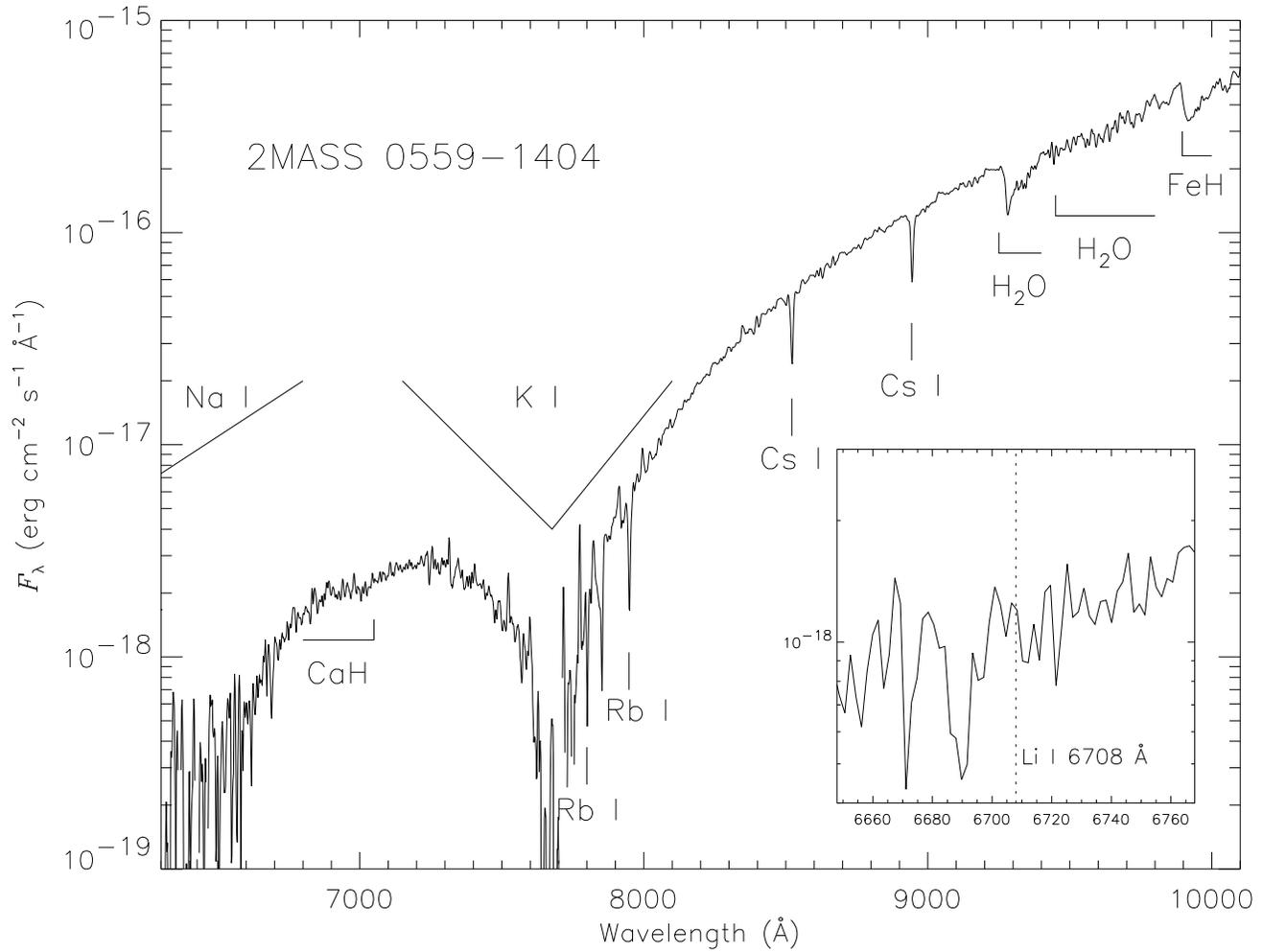}
\caption{Spectrum of 2MASS 0559$-$1404, scaled to the flux
density at 10 pc using I$_c$ and parallax measurements from \citet{dah02}.
Data are plotted on a logarithmic scale, and prominent spectral features are noted.
The inset box shows a close-up of the 6650--6770 {\AA} region around the
6708 {\AA} Li I line.  An offset feature is seen but is likely a noise spike;
no obvious Li I line is detected.
\label{fig2}}
\end{figure}

\begin{figure}
\epsscale{0.4}
\plotone{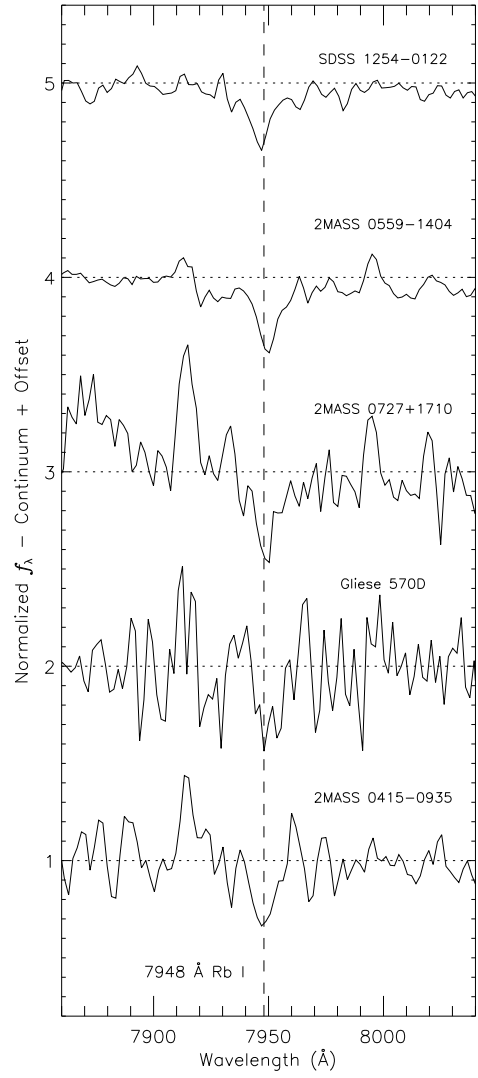}
\caption{7948 {\AA} Rb I line (dashed vertical line)
in the spectra of SDSS 1254$-$0122,
2MASS 0559$-$1404, 2MASS 0727+1710, Gliese 570D, and 2MASS 0415$-$0122.
Spectra are normalized between 7900 and 8000 {\AA}, and linear fits to the
local pseudo-continua have been subtracted.  Spectra are offset
by a constant for clarity (dotted lines).
Note that ``emission features'' at 7910 and 7990 {\AA}
are poorly-subtracted telluric OH skylines.
\label{fig3}}
\end{figure}

\begin{figure}
\epsscale{0.4}
\plotone{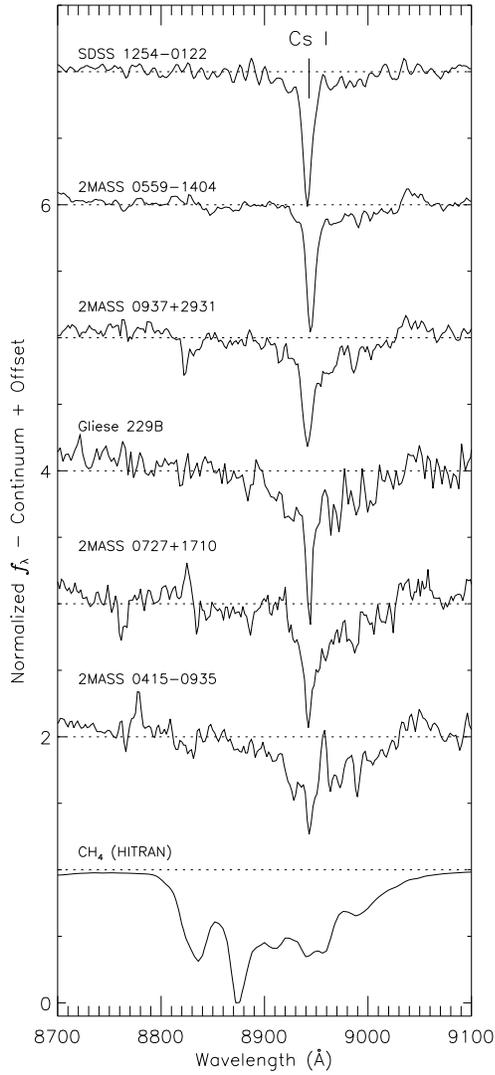}
\caption{Close-up of the 8700--9100 {\AA} region in the spectra of
SDSS 1254$-$0122, 2MASS 0559$-$1404, 2MASS 0937+2931, Gliese 229B
(data from Oppenheimer et al.\ 1998),
2MASS 0727+1710, and 2MASS 0415$-$0935.  Spectra are normalized as in Figure 3.
A normalized CH$_4$ opacity spectrum generated from data in the HITRAN database
is plotted at the
bottom for comparison.  The 8943 {\AA} Cs I line is
indicated, and Lorentzian profile fits to the lines in each spectrum
are delineated by dashed lines.
\label{fig4}}
\end{figure}

\begin{figure}
\epsscale{0.4}
\plotone{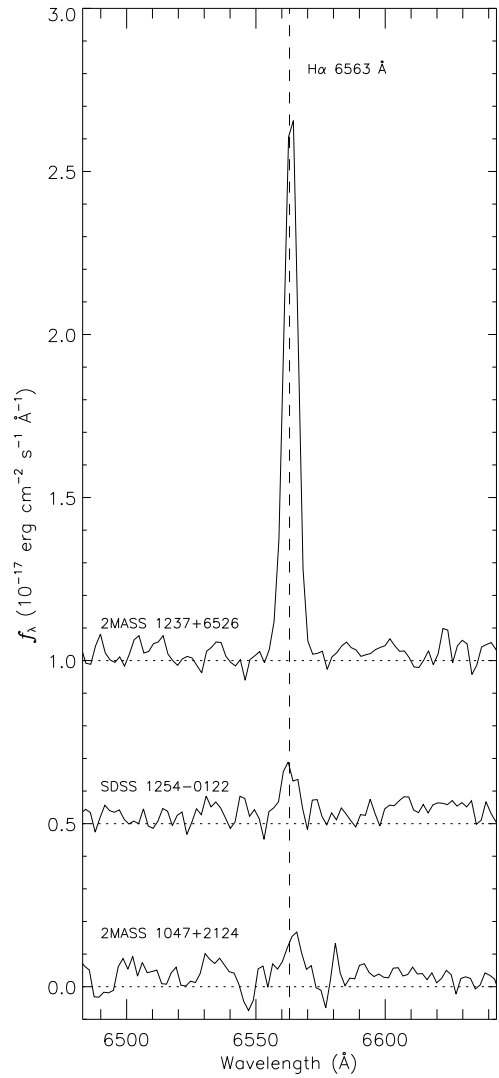}
\caption{Spectral region around the 6563 {\AA} H$\alpha$ emission line
(vertical dashed line) for 2MASS 1237+6526 (data from 2001 February 20 UT),
SDSS 1254$-$0122, and 2MASS 1047+2124.  Spectra are offset by a constant for
clarity (dotted line).
\label{fig5}}
\end{figure}

\begin{figure}
\epsscale{1.1}
\plottwo{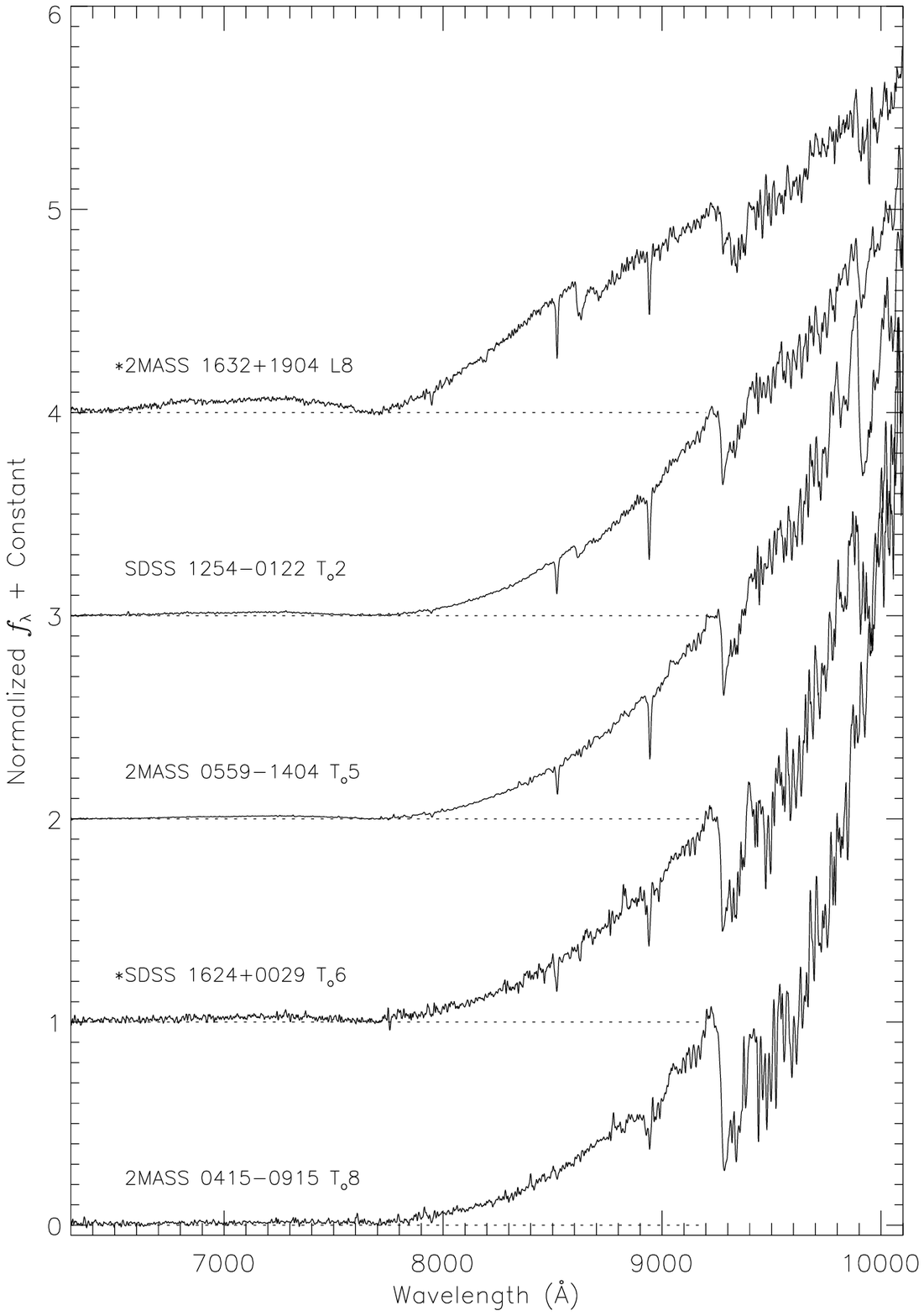}{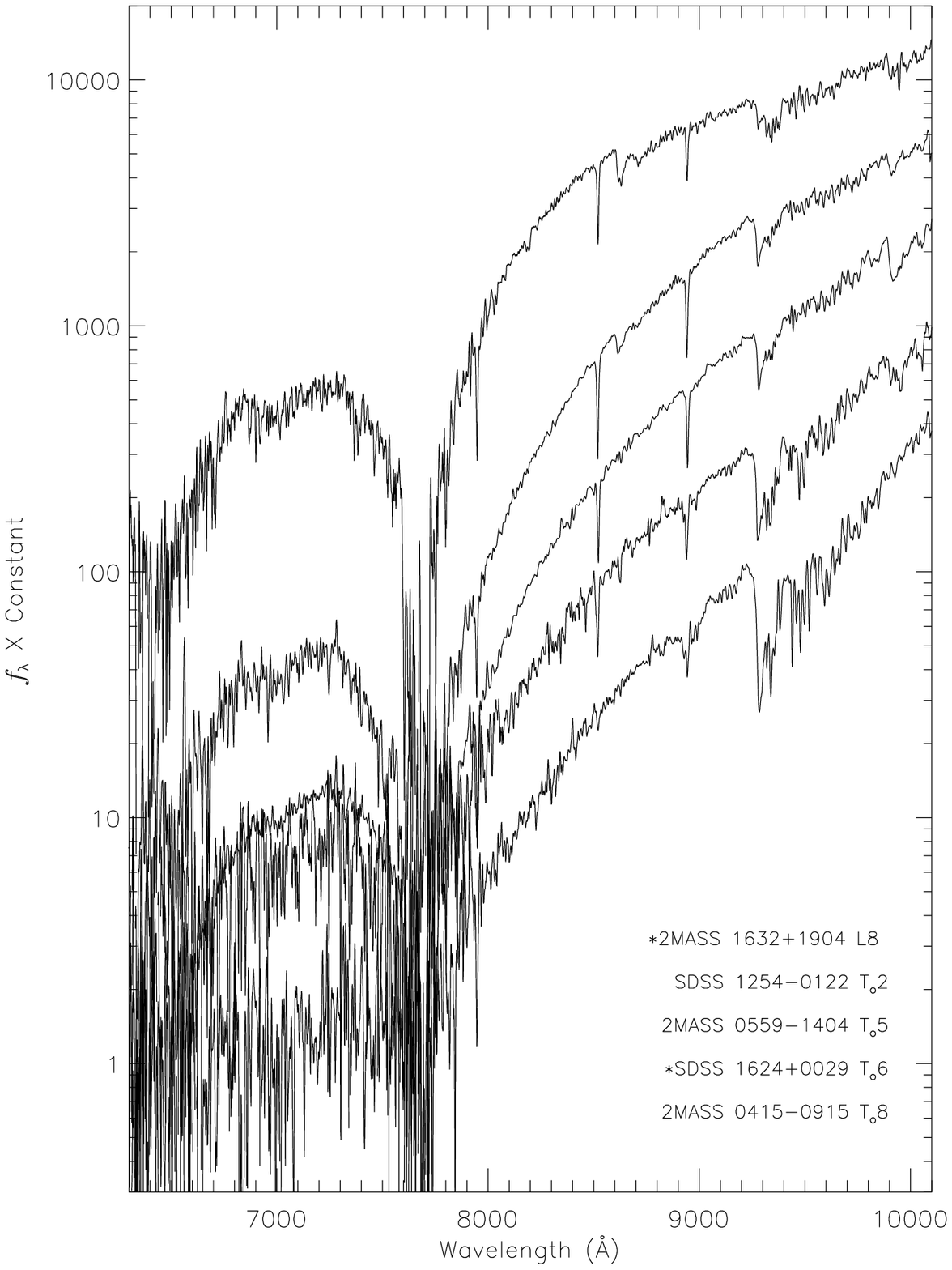}
\caption{LRIS spectra of optical spectral standards
2MASS 1632+1904 (L8),
SDSS 1254$-$0122 (T$_o$2), 2MASS 0559$-$1404 (T$_o$5),
SDSS 1624+0029 (T$_o$6), and 2MASS 0415$-$0935 (T$_o$8).
Spectra are normalized at 9250 {\AA}.  Objects labelled by an asterisk have not
had their spectra corrected for telluric absorption.
The left panel shows the standard spectra on a linear scale,
offset by a constant; the right panel shows the spectra on a
logarithmic scale, offset by a constant factor. \label{fig6}}
\end{figure}

\begin{figure}
\epsscale{0.8}
\plotone{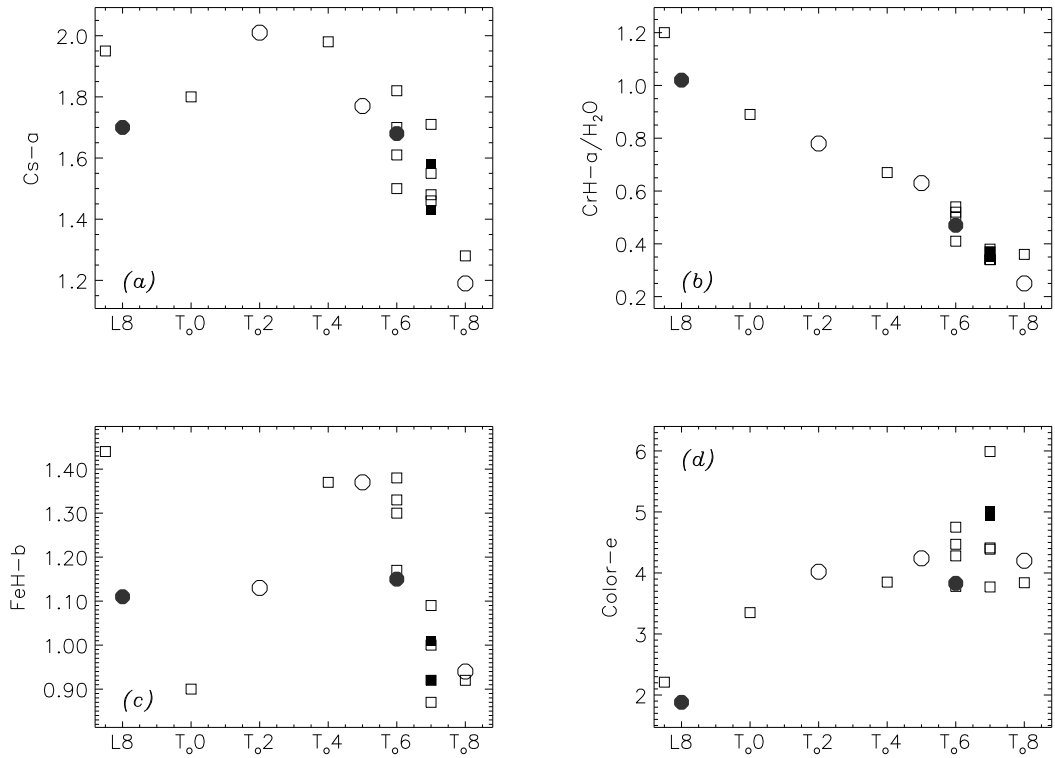}
\caption{Spectral indices versus final optical spectral type (Table 7, column 6):
(a) CsI(A), (b) CrH(A)/H$_2$O,
(c) FeH(B), and (d) Color-e.  Spectral standards are indicated by
circles, other spectra by squares.  Spectra that have not been corrected for telluric absorption
are indicated by filled symbols. Spectral type uncertainties are nominally $\pm$1 subclass, except
for SDSS 0837$-$0000 and
SDSS 1021$-$0304 which have explicitly labelled error bars. \label{fig7}}
\end{figure}

\begin{figure}
\epsscale{0.9}
\plotone{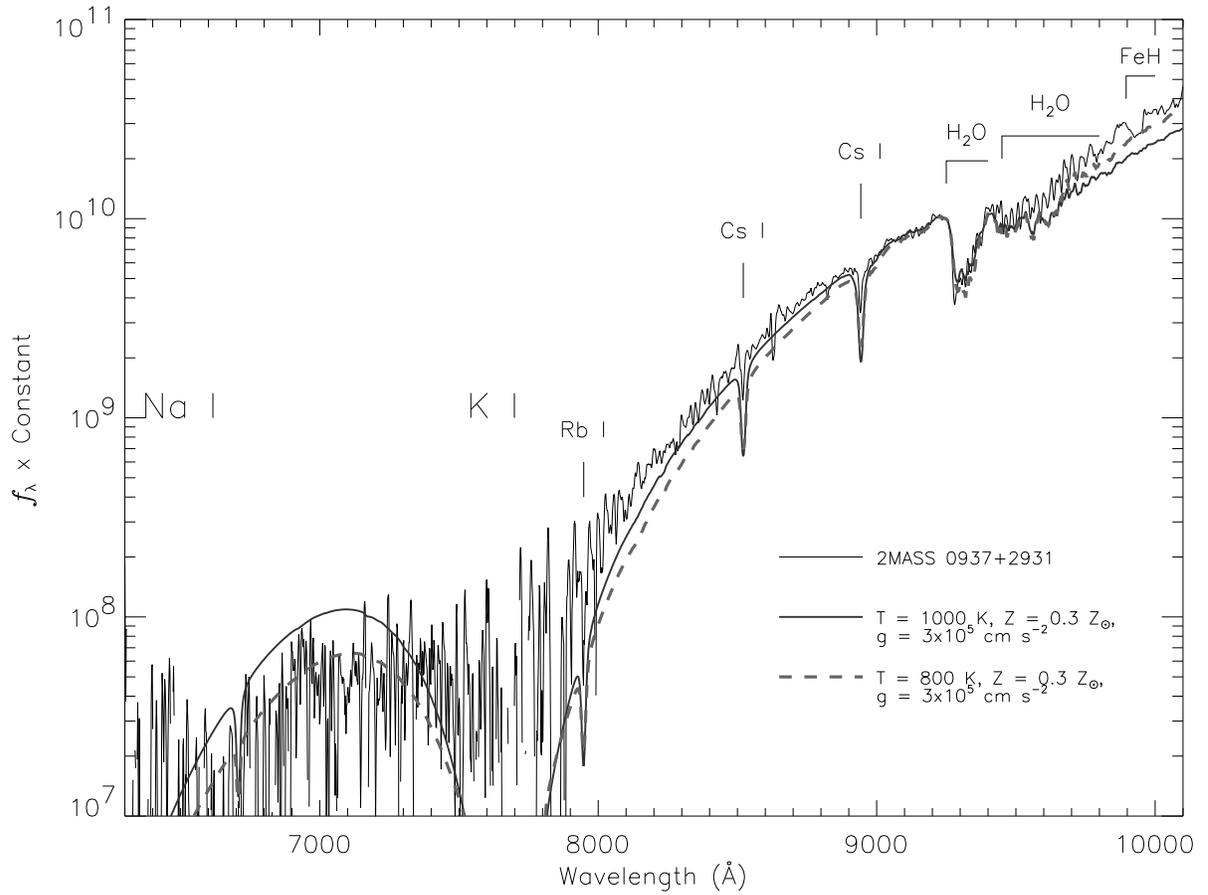}
\caption{Spectral model fits to the peculiar T dwarf 2MASS 0937+2931.
The best-fitting models from \citet{bur02} are shown, both with
g = $3{\times}10^5$ cm s$^{-2}$ and $Z = 0.3 Z_{\sun}$, and having
T$_{eff}$ = 1000 (solid grey line) and 800 K (dashed grey line).
Higher metallicity and/or lower gravity models do not match either
the steep spectral slope over 8000--10000 {\AA} or the suppression
of flux at $\sim$7100 {\AA} between the broadened Na I and K I lines.
Parallax measurement may provide a better constraint on the
T$_{eff}$ of this object. \label{fig8}}
\end{figure}

\begin{figure}
\epsscale{0.9}
\plotone{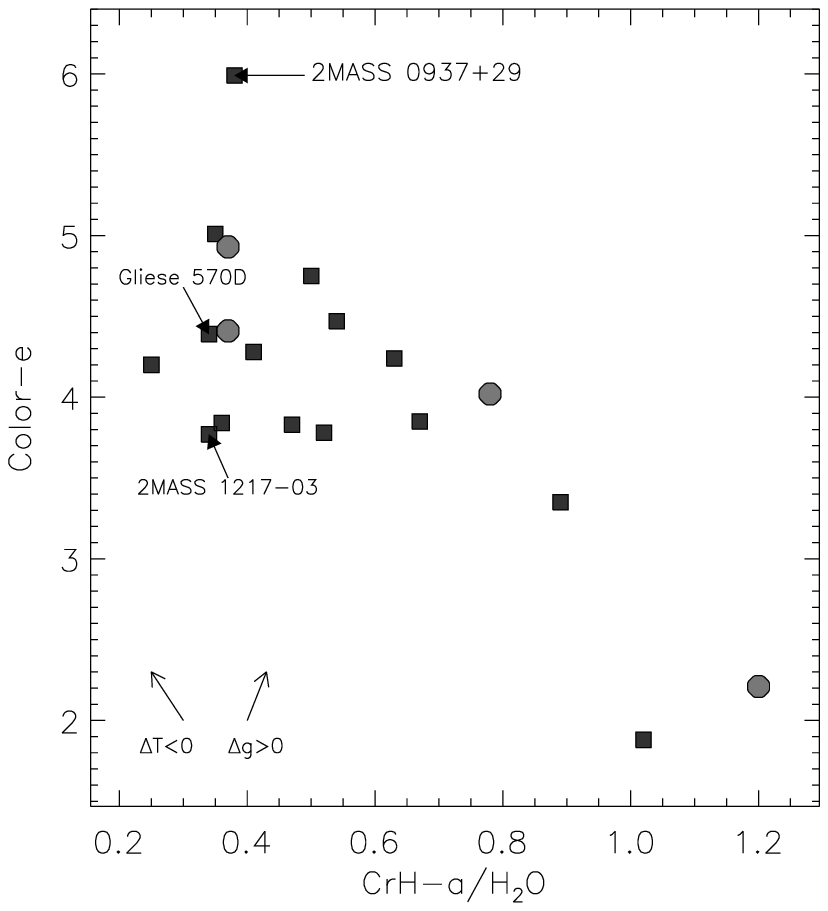}
\caption{Color-e versus CrH(A)/H$_2$O for our sample.  Objects
with H$\alpha$ emission are indicated by circles,
all others are indicated by squares.  Three objects are labelled:
the metal-poor and probable high-gravity
T dwarf 2MASS 0937+2931, the moderate-gravity companion T dwarf Gliese 570D
\citep[$\tau$ = 2--5 Gyr, g = (1--2)${\times}10^5$ cm s$^{-2}$]{me00a,geb01},
and the possible low-gravity T dwarf 2MASS 1217$-$0311 \citep{leg03}.
General dependencies on decreasing temperature and increasing
gravity at T$_{eff}$ = 1000 K and $Z = Z_{\sun}$, based on the models of \citet{bur02},
are indicated by the arrows in the lower
left corner. \label{fig9}}
\end{figure}


\begin{thebibliography}{}

\bibitem[Ackerman \& Marley(2001)]{ack01}Ackerman, A.\ S., \& Marley, M.\ S.
2001, \apj, 556, 872

\bibitem[Allard et al.(2001)]{all01}Allard, F., Hauschildt, P.\ H.,
Alexander, D.\ R., Tamanai, A., \&  Schweitzer, A. 2001, \apj, 556, 357

\bibitem[Allard et al.(1996)]{all96}Allard, F., Hauschildt, P.\ H.,
Baraffe, I., \& Chabrier, G. 1996, \apj, 465, L123

\bibitem[Anders \& Grevesse(1989)]{and89}Anders, E., \& Grevesse, N. 1989,
Geo.\ Cosmo.\ Acta, 53, 197

\bibitem[Auman(1967)]{aum67}Auman, J., Jr. 1967, \apjs, 14, 171

\bibitem[Berg \& Klynning(1974)]{ber74}Berg, L.\ E., \& Klynning, L. 1974, \aaps, 13, 325

\bibitem[Borysow, J{\o}rgensen, \& Zheng(1997)]{bor97}Borysow, A., J{\o}rgensen,
U.\ G., \& Zheng, C. 1997, \aap, 324, 185

\bibitem[Burgasser(2001)]{me01}Burgasser, A.\ J. 2001, Ph.D.\ Thesis,
California Institute of Technology

\bibitem[Burgasser et al.(2003a)]{me03a}Burgasser, A.\ J., Kirkpatrick,
J.\ D., McElwain, M.\ W., Cutri, R.\ M., Burgasser, A.\ J., \& Skrutskie, M.\ F.
2003a, \aj, 125, 850

\bibitem[Burgasser et al.(2003b)]{me03b}Burgasser, A.\ J., Kirkpatrick,
J.\ D., Reid, I.\ N., Brown, M.\ E., Miskey, C.\ L., \& Gizis, J.\ E. 2003b,
\apj, 586, 512

\bibitem[Burgasser et al.(2000a)]{me00b}Burgasser, A.\ J., Kirkpatrick, J.\ D.,
Reid, I.\ N., Liebert, J., Gizis, J.\ E., \& Brown, M.\ E. 2000a, \aj, 120,
473

\bibitem[Burgasser et al.(2002a)]{me02b}Burgasser, A.\ J., Liebert, J.,
Kirkpatrick, J.\ D., \& Gizis, J.\ E. 2002a, \aj, 123, 2744

\bibitem[Burgasser et al.(2002b)]{me02c}Burgasser, A.\ J., Marley, M.\ S.,
Ackerman, A.\ S., Saumon, D.,
Lodders, K., Dahn, C.\ C., Harris, H.\ C., \& Kirkpatrick, J.\ D.
2002b, \apj, 571, L151

\bibitem[Burgasser et al.(1999)]{me99} Burgasser, A.\ J., et al. 1999, \apj, 522, L65

\bibitem[Burgasser et al.(2000b)]{me00a} ---. 2000b, \apj, 531, L57

\bibitem[Burgasser et al.(2000c)]{me00c} ---. 2000c, \aj, 120, 1100

\bibitem[Burgasser et al.(2002c)]{me02a} ---. 2002c, \apj, 564, 421

\bibitem[Burgasser et al.(2003c)]{me03c} ---. 2003c, \apj, in press (astro-ph/0304174)

\bibitem[Burrows et al.(2002a)]{bur02}Burrows, A., Burgasser, A.\ J., Kirkpatrick,
J.\ D., Liebert, J., Milsom, J.\ A., Sudarsky, D., \& Hubeny, I. 2002a, \apj, 573, 394

\bibitem[Burrows et al.(2002b)]{bur02b}Burrows, A., Ram, R.\ S., Bernath, P.,
Sharp, C.\ M., Milsom, J.\ A. 2002b, \apj, 577, 986

\bibitem[Burrows, Marley, \& Sharp(2000)]{bur00}Burrows, A., Marley, M.\ S.,
\& Sharp, C.\ M. 2000, \apj, 531, 438

\bibitem[Burrows \& Sharp(1999)]{bur99}Burrows, A., \& Sharp, C.\ M. 1999, \apj,
512, 843

\bibitem[Burrows \& Volobuyev(2003)]{bur03} Burrows, A., \& Volobuyev, M. 2003, \apj,
583, 985

\bibitem[Burrows et al.(1997)]{bur97}Burrows, A., et al.\ 1997, \apj, 491, 856

\bibitem[Chabrier \& Baraffe(1997)]{cha97}Chabrier, G., \& Baraffe, I.
1997, \aap, 327, 1039

\bibitem[Cushing et al.(2003)]{cus03}Cushing, M.\ C., Rayner, J.\ T.,
Davis, S.\ P., \& Vacca, W.\ D. 2003, \apj, 582, 1066

\bibitem[Dahn et al.(2002)]{dah02}Dahn, C.\ C., et al. 2002, \aj,
124, 1170

\bibitem[Dick \& Fink(1977)]{dic77}Dick, K.\ A., \& Fink, U. 1977, \jqsrt,
18, 433

\bibitem[Dulick et al.(2003)]{dul03} Dulick, M., Bauschlicher, C.\ W., Jr., Burrows, A.,
Sharp, C.\ M., Ram, R.\ S., \& Bernath, P. 2003, \apj, submitted

\bibitem[Enoch, Brown, \& Burgasser(2003)]{eno03}Enoch, M.\ L., Brown, M.\ E.,
\& Burgasser, A.\ J. 2003, \aj, in press

\bibitem[Epchtein et al.(1997)]{epc97}Epchtein, N., et al. 1997,
The Messenger, 87, 27

\bibitem[Geballe et al.(2001)]{geb01}Geballe, T.\ R., Saumon, D., Leggett,
S.\ K., Knapp, G.\ R., Marley, M.\ S., \& Lodders, K. 2001, \apj, 556, 373

\bibitem[Geballe et al.(2002)]{geb02}Geballe, T.\ R., et al. 2002, \apj, 564, 466

\bibitem[Gelino et al.(2002)]{gel02} Gelino, C.\ R., Marley, M.\
S., Holtzman, J.\ A., Ackerman, A.\ S., \& Lodders, K. 2002, \apj,
577, 433

\bibitem[Gizis(1997)]{giz97}Gizis, J.\ E. 1997, \aj, 113, 806

\bibitem[Gizis et al.(2000)]{giz00}Gizis, J.\ E., Monet, D.\ G., Reid, I.\ N.,
Kirkpatrick, J.\ D., Liebert, J., \& Williams, R. 2000, \aj, 120, 1085

\bibitem[Golimowski et al.(1998)]{gol98}Golimowski, D.\ A., Burrows, C.\ S., Kulkarni, S.\ R.,
Oppenheimer, B.\ R., \& Brukardt, R.\ A. 1998, \aj, 115, 2579

\bibitem[Griffith \& Yelle(1999)]{gri99}Griffith, C.\ A., \& Yelle, R.\ V.
1999, \apj, 519, L85

\bibitem[Hall(2002a)]{hal02a}Hall, P.\ B. 2002a, \apj, 564, L89

\bibitem[Hall(2002b)]{hal02b}---. 2002b, \apj, 580, L77

\bibitem[Hamuy et al.(1994)]{ham94}Hamuy, M., Suntzeff, N.\ B., Heathcote, S.\ R., Walker, A.\ R.,
Gigoux, P., \& Phillips, M.\ M. 1994, PASP, 106, 566

\bibitem[Harris et al.(2003)]{har03}Harris, H.\ C., et al. 2003, \apj, in preparation

\bibitem[Hawley, Gizis, \& Reid(1996)]{haw96}Hawley, S.\ L., Gizis, J.\ E.,
\& Reid, I.\ N. 1996, \aj, 112, 2799

\bibitem[Kirkpatrick, Henry, \& McCarthy(1991)]{kir91}Kirkpatrick, J.\ D.,
Henry, T.\ J., \& McCarthy, D.\ W., Jr. 1991, \apjs, 77, 417

\bibitem[Kirkpatrick et al.(2000)]{kir00}Kirkpatrick, J.\ D., Reid, I.\ N.,
Liebert, J., Gizis, J.\ E., Burgasser, A.\ J., Monet, D.\ G., Dahn, C.\ C.,
Nelson, B., \& Williams, R.\ J. 2000, \aj, 120, 447

\bibitem[Kirkpatrick et al.(1999)]{kir99}Kirkpatrick, J.\ D., et al. 1999,
\apj, 519, 802

\bibitem[Kirkpatrick et al.(2003)]{kir03} ---. 2003, \aj, in preparation

\bibitem[Kleman \& Uhler(1959)]{kle59} Kleman, B., \& Uhler, U. 1959, Can.\ J.\ Phys., 37, 537

\bibitem[Leggett et al.(2003)]{leg03} Leggett, S.\ K., Golimowski, D.\ A.,
Fan, X., Geballe, T.\ R., \& Knapp, G.\ R. 2003, in Proceedings of the
12th Cambridge Workshop on Cool Stars, Stellar Systems, and the Sun,
ed.\ A.\ Brown, T.\ R.\ Ayres, \& G.\ M.\ Harper (Boulder: Univ.\ Colorado Press),
p.\ 120

\bibitem[Leggett et al.(1999)]{leg99} Leggett, S.\ K., Toomey, D.\ W.,
Geballe, T.\ R., \& Brown, R.\ H. 1999, \apj, 517, L139

\bibitem[Leggett et al.(2000)]{leg00}Leggett, S.\ K., et al. 2000,
\apj, 536, L35

\bibitem[Liebert et al.(2003)]{lie03}Liebert, J., Kirkpatrick, J.\ D., Cruz, K.\ L,
Reid, I.\ N., Burgasser, A.\ J., Tinney, C.\ G., \& Gizis, J.\ E.
2003, \aj, 125, 343

\bibitem[Liebert et al.(2000)]{lie00}Liebert, J., Reid, I.\ N., Burrows, A.,
Burgasser, A.\ J., Kirkpatrick, J.\ D., \& Gizis, J.\ E. 2000, \apj, 533,
L155

\bibitem[Liebert, Wehrse, \& Green(1987)]{lie87}Liebert, J., Wehrse, R., \&
Green, R.\ F. 1987, \aap, 175, 173

\bibitem[Lodders(1999)]{lod99}Lodders, K. 1999, \apj, 519, 793

\bibitem[Marley et al.(2002)]{mar02}Marley, M.\ S., Seager, S., Saumon, D.,
Lodders, K., Ackerman, A.\ S., Freedman, R., \& Fan, X. 2002, \apj, 568, 335

\bibitem[Mart{\'{\i}}n et al.(1999)]{mrt99}Mart{\'{\i}}n, E.\ L., Delfosee, X.,
Basri, G., Goldman, B., Forveille, T., \& Zapatero Osorio, M.\ R. 1999, \aj, 118, 2466

\bibitem[Mart{\'{\i}}n, Rebolo, \& Zapatero Osorio(1996)]{mrt96}Martin, E.\ L.,
Rebolo, R., \& Zapatero Osorio, M.\ R. 1996, \apj, 469, 706

\bibitem[McCook \& Sion(1999)]{mco99}McCook, G.\ P., \& Sion, E.\ M. 1999,
\apjs, 121, 1

\bibitem[Mohanty et al.(2002)]{moh02}Mohanty, S., Basri, G., Shu, F., Allard, F., \& Chabrier, G. 2002,
\apj, 572, 469

\bibitem[Morgan \& Keenan(1973)]{mor73} Morgan, W.\ W., \&
Keenan, P.\ C. 1973, \araa, 11, 29

\bibitem[Nakajima et al.(1995)]{nak95}Nakajima, T., Oppenheimer, B.\ R.,
Kulkarni, S.\ R.,
Golimowski, D.\ A., Matthews, K., \& Durrance, S.\ T. 1995, \nat, 378, 463

\bibitem[Nakajima et al.(2000)]{nak00}Nakajima, T., Tsuji, T., Tamura, M., \&
Yamashita, T. 2000, PASJ, 52, 87

\bibitem[Nakajima, Tsuji, \& Yanagisawa(2001)]{nak01}Nakajima, T., Tsuji, T., \& Yanagisawa, K.
2001, \apj, 561, L119

\bibitem[Oke et al.(1995)]{oke95}Oke, J.\ B., et al. 1995, \pasp, 107, 375

\bibitem[Oppenheimer et al.(1998)]{opp98}Oppenheimer, B.\ R., Kulkarni, S.\ R.,
Matthews, K., van Kerkwijk, M.\ H. 1998, \apj, 502, 932

\bibitem[Oppenheimer et al.(1995)]{opp95}Oppenheimer, B.\ R., Kulkarni, S.\ R.,
Matthews, K., \& Nakajima, T. 1995, Science, 270, 1478

\bibitem[Putney(1997)]{put97} Putney, A. 1997, \apjs, 112, 527

\bibitem[Reid et al.(2001)]{rei01}Reid, I.\ N., Burgasser, A.\ J.,
Cruz, K., Kirkpatrick, J.\ D., \& Gizis, J.\ E. 2001, \aj, 121, 1710

\bibitem[Reid \& Hawley(2000)]{rei00b}Reid, I.\ N., \& Hawley, S.\ L.
2000, New Light on Dark Stars (Chichester: Praxis)

\bibitem[Rothman et al.(1998)]{rot98}Rothman, L.\ S., et al. 1998, JQSRT, 60, 665

\bibitem[Shultz et al.(1998)]{sch98} Shultz, A.\ B., et al. 1998, \apj, 492, L181

\bibitem[Skrutskie et al.(1997)]{skr97}Skrutskie, M.\ F., et al. 1997, in The Impact of Large-Scale Near-IR
Sky Surveys, ed. F.\ Garzon (Dordrecht: Kluwer), p.\ 25

\bibitem[Stevenson(1994)]{ste94}Stevenson, C.\ C. 1994, \mnras, 267, 904

\bibitem[Tsuji(2002)]{tsu02}Tsuji, T. 2002, \apj, 575, 264

\bibitem[Tsuji, Ohnaka, \& Aoki(1996)]{tsu96}Tsuji, T., Ohnaka, K., \&
Aoki, W. 1996, \aap, 308, L29

\bibitem[Tsuji, Ohnaka, \& Aoki(1999)]{tsu99} Tsuji, T., Ohnaka, K., \&
Aoki, W. 1999, \apj, 520, L119

\bibitem[Wiese, Smith, \& Glennon(1966)]{wie66}Wiese, W.\ L., Smith, M.\ W.,
\& Glennon, B.\ M. 1966, Atomic Transition Probabilities, Vol.\ 1.
(Washington, D.\ C.: GPO)

\bibitem[Wing, Cohen, \& Brault(1977)]{win77} Wing, R.\ F., Cohen, J., \& Brault,
J.\ W. 1977, \apj, 216, 659

\bibitem[York et al.(2000)]{yor00} York, D.\ G., et al. 2000, AJ,
120, 1579

\end{thebibliography}
\end{document}